\documentclass[acmsmall, screen]{acmart}

\usepackage{algorithm}
\usepackage{algorithmic}
\usepackage{subfigure}
\usepackage{xcolor}
\usepackage{wrapfig}
\usepackage{enumitem}
\newcommand{\scheme}{EfficientUICoder}
\newcommand{\Function}[2]{\STATE \textbf{Function} #1(#2)}

\definecolor{bg}{HTML}{F8F9FB}  
\definecolor{rowcolor}{HTML}{ECEFF4}
\definecolor{greenbg}{HTML}{D2E4DB}
\definecolor{added}{HTML}{C7E0D6}
\definecolor{removed}{HTML}{FBDBD8}

\newcommand{\revision}[1]{\textcolor{black}{#1}}

\usepackage{tcolorbox}
\usepackage{multirow}

\AtBeginDocument{%
  }

\begin{document}




\title[\scheme: A Bidirectional Token Compression Framework for Efficient MLLM-Based UI Code...]{\scheme: A Bidirectional Token Compression Framework for Efficient MLLM-Based UI Code Generation}

\author{Jingyu Xiao}
\orcid{0000-0002-2394-2995}
\affiliation{%
  \institution{The Chinese University of Hong Kong}
  \city{Hong Kong}
  \country{China}
}
\email{jyxiao@link.cuhk.edu.hk}

\author{Zhongyi Zhang}
\orcid{0009-0009-9951-3335}
\affiliation{%
  \institution{Huazhong University of Science and Technology}
  \city{Wuhan}
  \country{China}}
\email{zyzhang1@hust.edu.cn}

\author{Yuxuan Wan}
\orcid{0009-0006-6739-4675}
\affiliation{%
  \institution{The Chinese University of Hong Kong}
  \city{Hong Kong}
  \country{China}
}
\email{yxwan9@cse.cuhk.edu.hk}

\author{Yintong Huo}
\orcid{0009-0006-8798-5667}
\authornote{Yintong Huo is the corresponding author.}
\affiliation{%
 \institution{Singapore Management University}
 \city{Singapore}
 \country{Singapore}}
\email{ythuo@smu.edu.sg}

\author{Yang Liu}
\orcid{0000-0001-7300-9215}
\affiliation{%
  \institution{Nanyang Technological University}
  \city{Singapore}
  \country{Singapore}}
\email{yangliu@ntu.edu.sg}

\author{Michael R. Lyu}
\orcid{0000-0002-3666-5798}
\affiliation{%
  \institution{The Chinese University of Hong Kong}
  \city{Hong Kong}
  \country{China}}
\email{lyu@cse.cuhk.edu.hk}

\begin{abstract}

Multimodal Large Language Models (MLLMs) have demonstrated exceptional performance in UI2Code tasks (i.e., generating code from UI mockups), significantly enhancing website development efficiency. However, UI2Code tasks incur substantially higher computational overhead compared to traditional code generation tasks. This overhead is primarily driven by the large number of input image tokens required to represent complex visual designs and the extensive volume of output code tokens needed to describe complete webpage structures. In this paper, we conduct a comprehensive preliminary study on popular MLLMs for UI2Code tasks, identifying significant redundancies in both image and code tokens. We observe that these redundancies not only exacerbate computational complexity but also hinder the model’s ability to focus on key UI elements, leading to excessively lengthy and often invalid HTML files. To address these challenges, we propose \scheme, a bidirectional compression framework designed for efficient UI code generation. First, we introduce an Element and Layout-aware Token Compression method, which preserves essential UI element and layout information by detecting element regions and constructing a UI element tree for efficient representation. Second, we design a Region-aware Token Refinement strategy that refines selected tokens by leveraging attention scores to evaluate semantic importance, discarding low-attention tokens from selected regions while integrating high-attention tokens from unselected regions. Third, we develop an Adaptive Duplicate Token Suppression mechanism, which dynamically modulates token probabilities during decoding by tracking HTML/CSS code structure frequencies and applying exponential penalty strategies to minimize repetitive generation. Extensive experiments demonstrate that \scheme \ achieves a \textbf{55\%-60\%} compression ratio without compromising the quality of the generated webpages, effectively reducing output code redundancy. In terms of efficiency, \scheme \ achieves superior improvements, reducing computational cost by up to \textbf{44.9\%}, generated tokens by up to \textbf{41.4\%}, prefill time by up to \textbf{46.6\%}, and inference time by up to \textbf{48.8\%} on 34B-level MLLMs. Code is available at \url{https://github.com/WebPAI/EfficientUICoder}.

\end{abstract}


\setcopyright{cc}
\setcctype{by}
\acmJournal{PACMSE}
\acmYear{2026} \acmVolume{3} \acmNumber{FSE} \acmArticle{FSE107}
\acmMonth{7} 
\acmDOI{10.1145/3808114}
\acmSubmissionID{fse26mainb-p200-p}
\received{2026-02-24}
\received[accepted]{2026-03-24}





\maketitle

\section{Introduction}


Converting webpage designs into functional UI code represents a critical yet labor-intensive bottleneck in web development. Manual translation of UI designs into code is both time-consuming and requires specialized domain expertise, creating significant barriers to rapid web development. Multimodal Large Language Models (MLLMs), with their remarkable capabilities in visual-language understanding, show significant potential for visually rich code generation applications, such as creating UIs~\cite{wan2024automatically,zhang2025empowering}, slides~\cite{tang2025slidecoder}, and posters~\cite{pang2025paper2poster}.
This success has recently spurred considerable interest within the software engineering research community, particularly in UI-to-code (UI2Code) conversion.
For example, DCGen~\cite{wan2024automatically} employs a divide-and-conquer strategy to generate submodule code before assembly, while DeclarUI~\cite{zhou2024bridging} combines element segmentation with page transition graphs for mobile app UI generation. UICopilot~\cite{gui2025UICopilot} adopts hierarchical generation by creating HTML structures before components, and LayoutCoder~\cite{wu2024mllm} introduces layout-aware frameworks for preserving UI layout fidelity. However, none of them consider the computational overhead and efficiency of UI2Code tasks.



\begin{figure}[h]
\includegraphics[width = .8\textwidth]{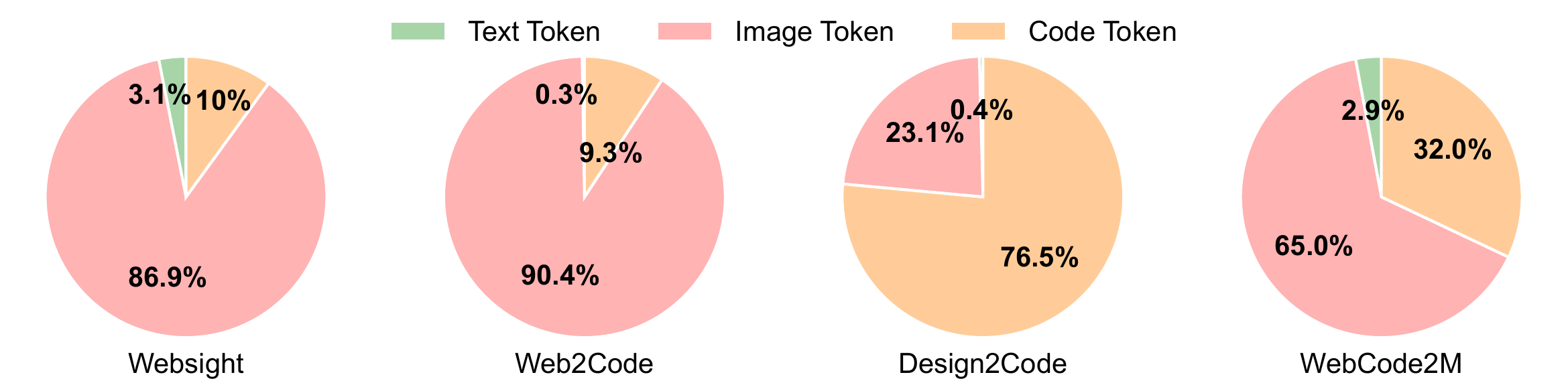}
\caption{The token ratios of different datasets.}
\label{fig:ratio}
\end{figure}

Compared to traditional code generation tasks~\cite{docstring2025yang, pan2025hidden, sun2024tokensugar,shi2025efficient}, UI2Code presents unique computational challenges. These tasks consume substantially more tokens due to two primary factors: the extensive number of input image tokens required to represent complex visual designs, and the large volume of generated code tokens needed to describe complete webpage structures. We analyze the token distribution across several mainstream UI2Code benchmarks, including WebSight~\cite{laurençon2024unlocking}, Web2code~\cite{yun2024web2code}, Design2Code~\cite{si2025design2code}, and WebCode2M~\cite{gui2025webcode2m}. As illustrated in Fig.~\ref{fig:ratio}, the combined image token and code token ratio accounts for more than 90\% of the total token consumption.


These excessively long sequences of image and code tokens consume substantial memory and computational resources throughout the entire MLLM pipeline during code generation. On the one hand, previous studies~\cite{dosovitskiy2020image} have demonstrated that the information contained in images is much sparser than that in text. On the other hand, several studies~\cite{si2025design2code, guo2024stop} have pointed out that code generated by LLMs frequently includes redundant information. Hence, a natural question arises: \textbf{``Are all image and code tokens necessary for UI code generation?''}

To identify redundancy in image tokens (Section~\ref{subsec:redundant_visual}), we conduct an in-depth analysis of visual tokens generated by the widely adopted vision encoder CLIP~\cite{radford2021learning}. Attention score visualization reveals that redundant visual tokens not only inflate computational costs but also misdirect attention toward uninformative regions (e.g., background areas), thereby diminishing the model's focus on critical UI elements. To examine code redundancy (Section~\ref{subsec:redundant_code}), we evaluate two prominent UI2Code benchmarks Design2Code and WebCode2M using popular open-source MLLMs (Llava-v1.6-7b and Llava-v1.6-34b~\cite{liu2024llavanext}). Through systematic analysis of the outputs, we identify three redundancy types, demonstrating that MLLMs frequently produce superfluous HTML/CSS structures and textual content. This code redundancy not only increases computational overhead but also entraps models in cyclical generation patterns, potentially resulting in invalid HTML structures.

Based on these findings, we propose the first multimodal token compression framework for UI-to-code, namely \scheme. \scheme\ mitigates redundant image tokens during the encoding stage and suppresses redundant code tokens during the decoding stage. First, we propose an \textbf{Element and Layout-aware Token Compression (ELTC)} method to compress redundant tokens while preserving UI element and layout information. It initially employs a UI element detection algorithm to identify element regions and constructs a UI element graph to model the UI layout structure. Subsequently, a minimum spanning tree algorithm is applied to obtain the most efficient UI representation in the form of a UI element tree. Second, we devise a \textbf{Region-aware Token Refinement (RTR)} strategy to refine the tokens selected in the first stage. It leverages attention scores to evaluate the semantic importance of visual tokens and removes tokens with lower attention scores that were selected by the first stage. Recognizing that certain critical tokens may reside in background regions, it selectively incorporates high-attention tokens from these areas. By carefully balancing the dropping ratio and selection ratio, we achieve substantial token compression while preserving the most semantically important visual information across both foreground and background regions. Third, we design an \textbf{Adaptive Duplicate Token Suppression (ADTS)} method to modulate token probabilities during the decoding stage. It first employs HTML and CSS parsers to track the frequencies of HTML/CSS code structures and textual content. Then an exponential penalty strategy is applied to adjust the probabilities of subsequent tokens, thereby suppressing repetitive generation based on the frequency of repetitions.  We conduct experiments on widely used UI2Code benchmark Design2Code and WebCode2M using Llava-v1.6 series models. Extensive experiments demonstrate that \scheme \ achieves a 55\%-60\% compression ratio without compromising UI2Code performance, effectively reducing output code redundancy. In terms of efficiency, \scheme \ achieves superior improvements, reducing computational cost by up to 44.9\%, generated tokens by up to 41.4\%, prefill time by up to 46.6\%, and inference time by up to 48.8\% on 34B-level MLLMs. Our contributions are as follows:

\begin{itemize}[leftmargin=*]
    \item We conduct an in-depth analysis of token redundancy in UI2Code tasks, identifying two redundancy patterns in the typical UI-to-code generation process.

    \item We propose \scheme, the first multimodal bidirectional token compression framework that reduces visual redundancy by constructing a UI element tree to preserve key UI elements and layout structures during encoding, while mitigating code redundancy by decreasing the probabilities of repeated tokens during the decoding stage.

    
    \item We conduct extensive experiments across popular UI2Code benchmarks and state-of-the-art MLLMs, demonstrating that our approach achieves substantial token compression and efficiency while maintaining generated UI quality.    
\end{itemize}

\section{Background}

\subsection{UI-to-Code Task Definition}
The UI2Code task involves translating visual webpage designs into functional HTML+CSS code that accurately reproduces the original design. Let $I_0$ denote the input design image of a webpage, and let $C_0$ represent the ground-truth HTML+CSS code. Given the design image $I_0$, a Multimodal Large Language Model $M$ generates code $C_g = M(I_0)$, where the rendered output $I_g$ produced by executing $C_g$ should closely approximate $I_0$ in both structural layout and visual appearance.

\begin{figure}[ht]
\centering
\includegraphics[width = .75\textwidth]{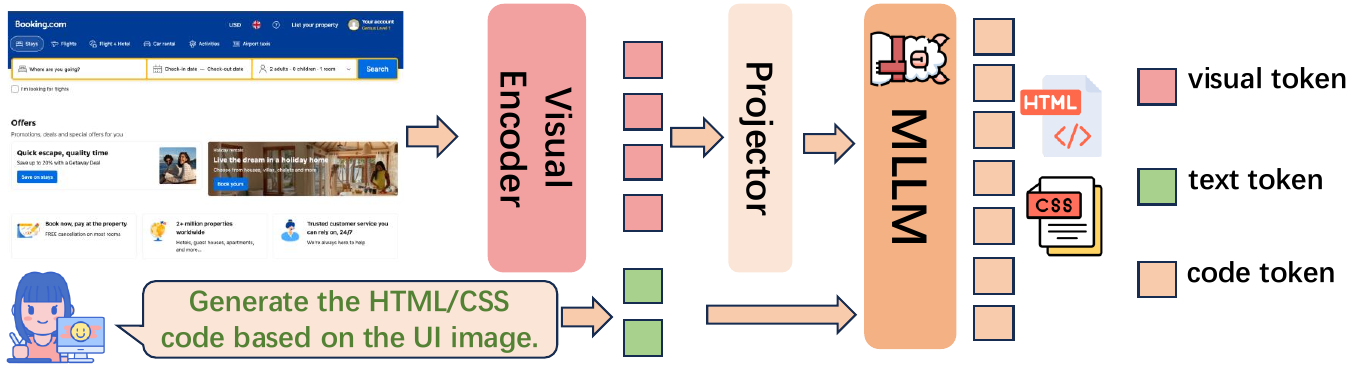}
\caption{MLLM-based UI2Code task pipeline.}
\label{fig:example}
\end{figure}

\subsection{MLLMs Background}

\textbf{Architecture.} As illustrated in Fig.~\ref{fig:example}, MLLM architectures typically comprise three core components: a visual encoder, a modality projector, and a large language model (LLM). The visual encoder, commonly implemented as a pre-trained image encoder such as CLIP's vision model~\cite{radford2021learning}, first converts the image into patch blocks and then transforms input images into visual token representations. The projector module serves as a bridge that aligns these visual tokens with the LLM's word embedding space, thereby enabling the LLM to effectively process visual information. Subsequently, the LLM integrates the aligned visual and textual representations to generate coherent responses.

\textbf{Computational Complexity.} Assessing the computational complexity of MLLMs necessitates analyzing key architectural components, particularly the self-attention mechanism and the feed-forward network (FFN). The total floating-point operations (FLOPs) can be formulated as:
\begin{equation} 
\text{Total FLOPs} = T \times \big( 4nd^2 + 2n^2d + 2ndm \big), 
\end{equation} 
where $T$ denotes the number of transformer layers, $n$ represents the sequence length, $d$ indicates the hidden dimension size, and $m$ corresponds to the intermediate size of the FFN. This formulation demonstrates that computational complexity exhibits a strong dependence on the sequence length $n$. In typical MLLM applications, the sequence length is defined as:
\begin{equation} 
n = n_{\text{img}} + n_{\text{text}}, 
\end{equation} 
where $n_{\text{img}}$ frequently exceeds $n_{\text{text}}$ by substantial margins, often by a factor of 20 or more. Thus, reducing $n_{\text{img}}$ represents a critical strategy for enhancing the computational efficiency of MLLMs.

\section{Preliminary Study}



\subsection{Visual Token Redundancy}
\label{subsec:redundant_visual}

In practical front-end development, developers don't need to examine every pixel to implement a webpage. Many visual components contain inherently redundant information, including image placeholder, uniform background regions and so on. For any given element, experienced programmers can infer critical specifications such as color, size, and styling properties from a subset of representative pixels rather than exhaustive full pixel visual analysis. 

\begin{figure}[ht]
    \subfigure[Search webpage and attention distribution on it.]{
    \label{fig:example1}
    \centering
    \includegraphics[width = .25\textwidth]{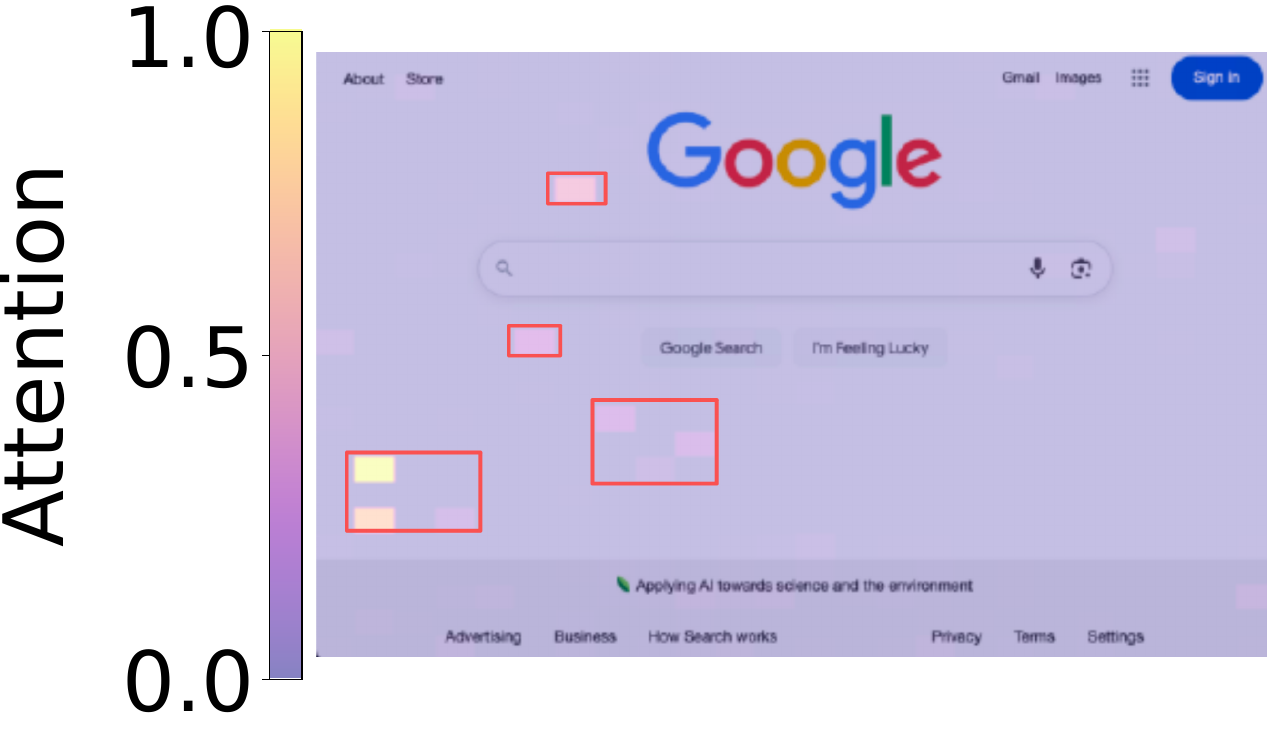}
    \includegraphics[width = .21\textwidth]{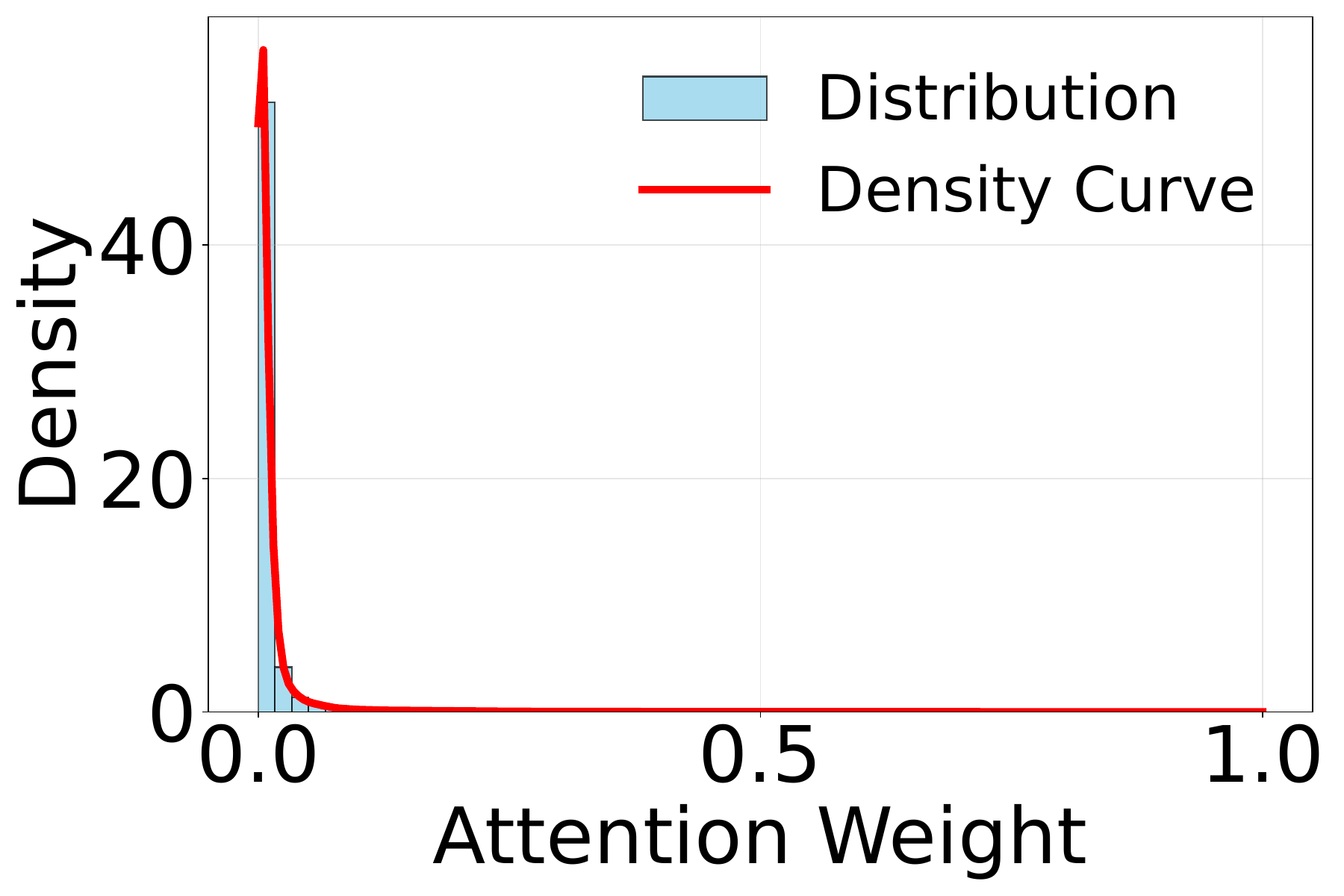}
    }    
    \subfigure[Shopping webpage and attention distribution on it.]{
    \label{fig:exampl2}
    \centering
    \includegraphics[width = .24\textwidth]{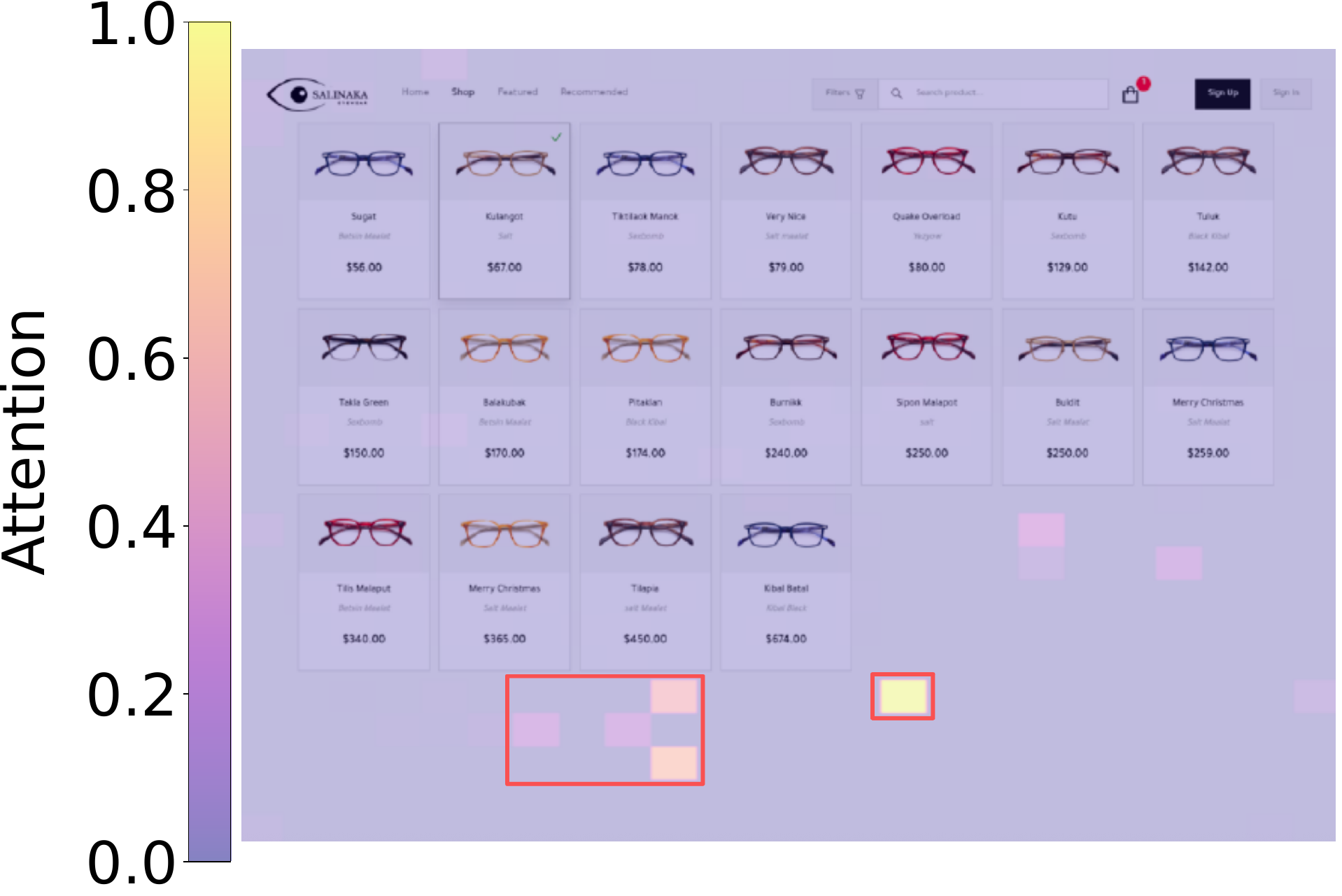}
    \includegraphics[width = .22\textwidth]{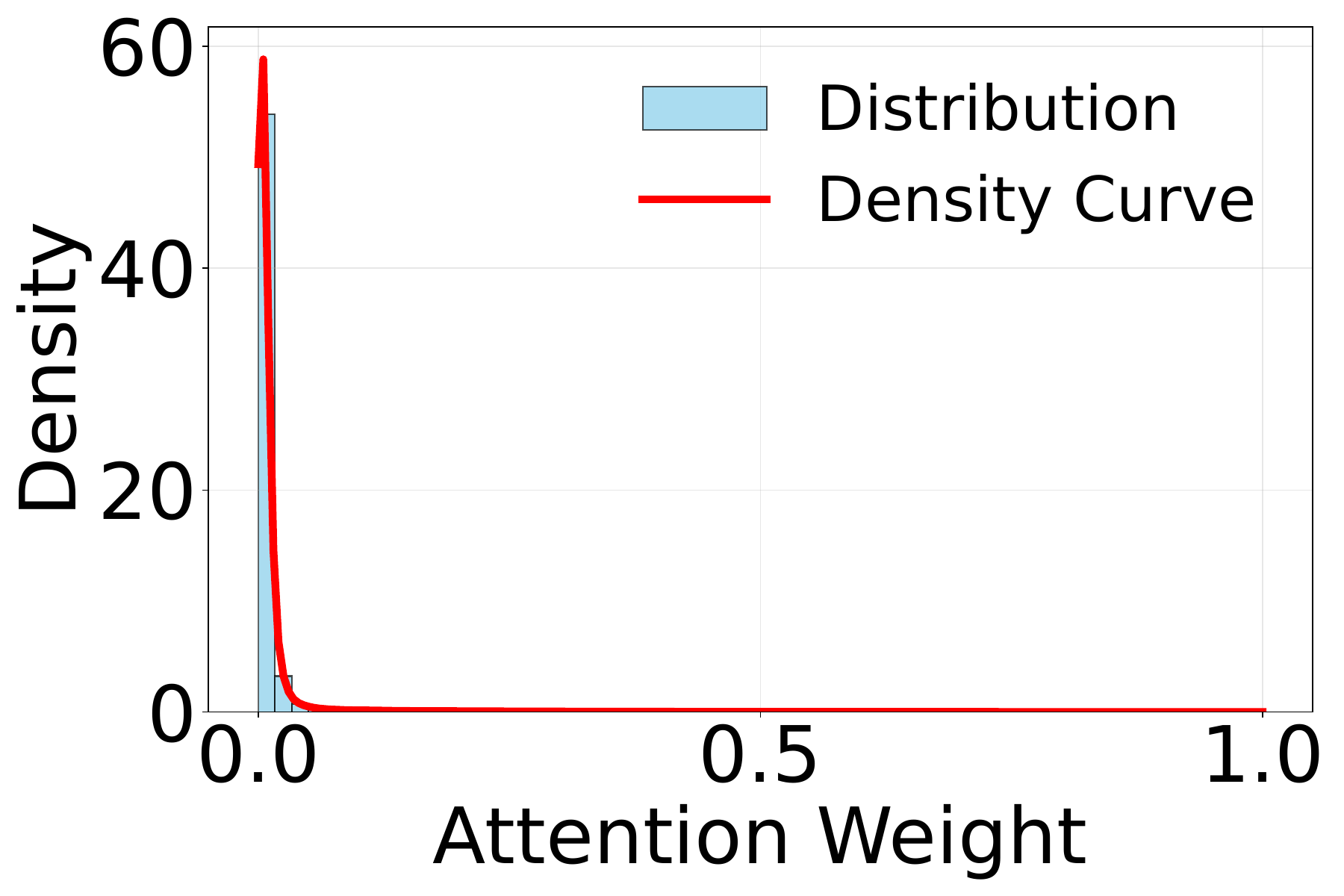}
    }
    \caption{Visual encoders' attention score visualization and distribution on two webpages.}
    \label{fig:visual_encoder}
\end{figure}

To investigate ``\textit{where are MLLMs looking at}'' on the UI image, we conduct an in-depth analysis on the visual tokens generated by the widely used vision encoders,
CLIP~\cite{radford2021learning}. Fig.~\ref{fig:visual_encoder} shows the normalized attention score distribution on two webpages (i.e., search engine webpage and shopping webpage), we use red bounding boxes to mark the area with high attention and find that only a few tokens receive high attention, while most visual tokens receive minimal attention. Surprisingly, the models pay more attention to the empty background part of the UI image than to the UI elements.

Redundant visual tokens introduce two detrimental effects on model performance: (1) \textit{Computational overhead escalation}: redundant visual tokens substantially increase the sequence length of inputs, leading to quadratic growth in attention computation complexity. (2) \textit{Attention distraction}: excessive redundant tokens (e.g., background) scatter the model's attention across uninformative visual regions, diminishing its capacity to focus on semantically crucial elements.

\begin{tcolorbox}[colback=gray!20, colframe=gray!20, width=\columnwidth, left=0.05in, right=0.05in, top=0.05in, bottom=0.05in]
\textbf{Observation 1:}  
Redundant visual tokens inflate computational cost and divert attention to uninformative regions (e.g., background), reducing the model’s focus on critical UI elements.  
\end{tcolorbox}

\subsection{Code Token Redundancy}
\label{subsec:redundant_code}

We conduct experiments on Llava-v1.6 (7B and 34B)~\cite{liu2024llavanext} to study their output code on UI2Code task. We select the widely used Design2Code~\cite{si2025design2code} dataset and adopt the ``direct prompt'' in design2code~\cite{si2025design2code} for evaluation. In our experiments, we deploy Multi-modal Large Language Models (MLLMs) on four NVIDIA RTX A800 GPUs, each with 80G memory. The maximum generated code length is set to 4096. And we set the temperatures as 0 for getting stable output.


\begin{figure}[t]
\centering
\includegraphics[width = .75\textwidth]{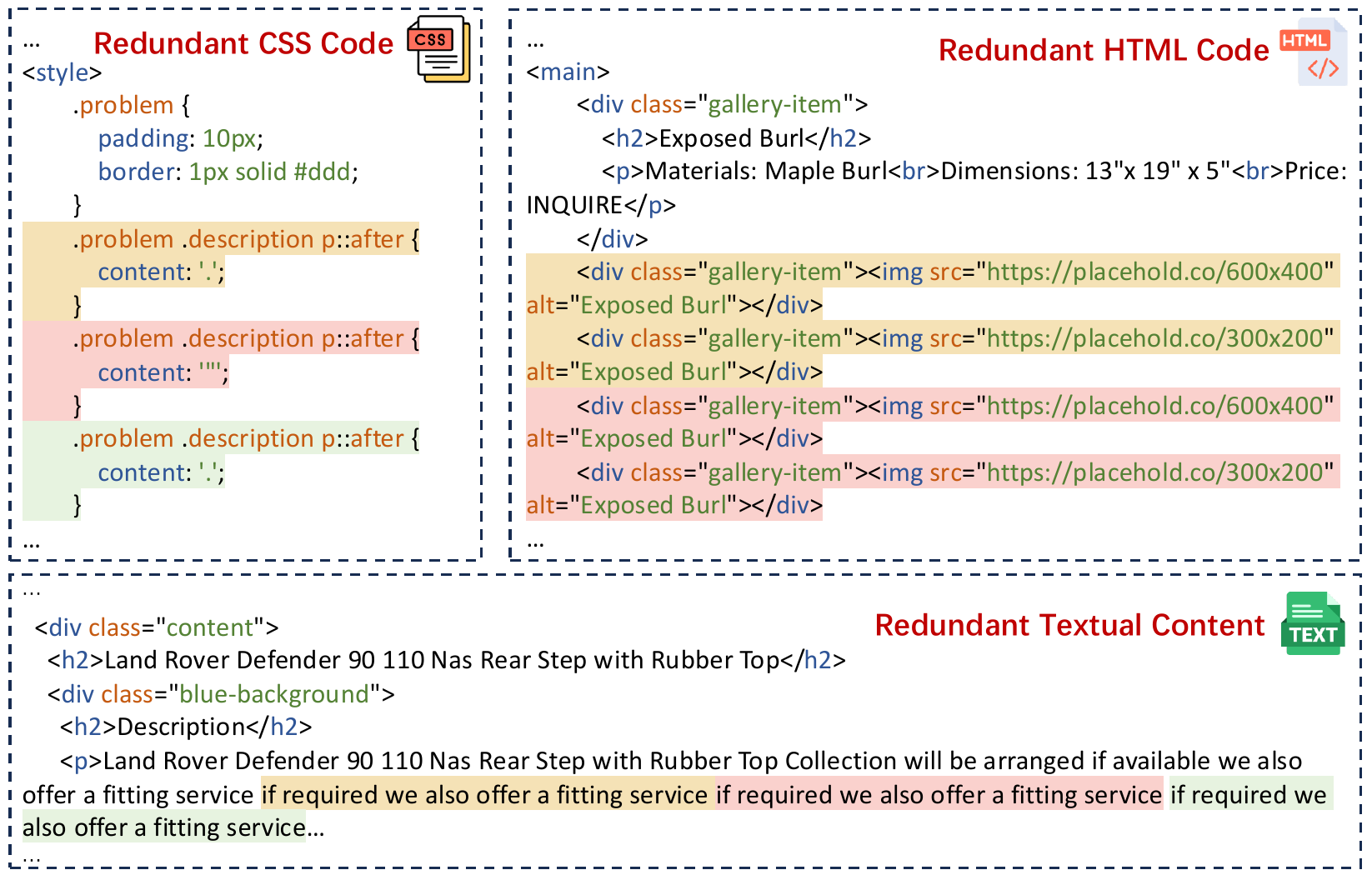}
\caption{Duplicate code token examples.}
\label{fig:dup_example}
\end{figure}


We engage a PhD student with extensive front-end development expertise to manually screen out samples with redundant content and then conduct analysis. Our systematic analysis reveals three primary categories of redundancy issues: (1) \textit{Redundant CSS code}: As demonstrated in the upper-left panel of Fig.~\ref{fig:dup_example}, identical CSS selectors such as ``.problem .description p::after'' are generated multiple times with duplicate styling properties. (2) \textit{Redundant HTML Code}: The upper-right panel illustrates unnecessary repetition of structural patterns, where multiple ``<div class="gallery-item">'' elements with identical content and attributes are generated redundantly. (3) \textit{Redundant textual content}: The bottom panel exemplifies duplicate text generation, where phrases such as ``we also offer a fitting service if required'' are repeated multiple times within the same content block.

This repetitive pattern may continue indefinitely until the model reaches its maximum token limit. We calculate the ratio of different issues, with results presented in Table~\ref{tab:code_redundancy}. Analysis of the rendered outputs reveals that such redundant generation introduces several detrimental effects: it inflates token consumption, increases webpage length, and extends generation time. More critically, excessive repetition can trap the model in generating a particular element, causing it to ignore subsequent elements or fail entirely to produce valid HTML structures.



\begin{table}[t]
\centering
\footnotesize
\caption{The redundant statistics on two datasets. ``Endless'' denotes the repetition continue until the model reaches its maximum token limit. ``End'' means that the repetition terminates after repeated several times.}
\label{tab:code_redundancy}
\resizebox{\textwidth}{!}{
\begin{tabular}{c|c|cc|cc|cc|c}
\hline
\multirow{2}{*}{Model} & \multirow{2}{*}{Dataset} & \multicolumn{2}{c|}{Redundant CSS Code} & \multicolumn{2}{c|}{Redundant HTML Code} & \multicolumn{2}{c|}{Redundant Textual Content} & \multirow{2}{*}{Total} \\
\cline{3-8}
 &  & End & Endless & End & Endless & End & Endless &  \\
\hline
\multirow{2}{*}{Llava-v1.6-7b} & Design2Code &  37 & 10  & 16  & 54  & 4  & 27  & 148 \\ 
 & WebCode2M   & 7 & 1 & 5 & 19 & 0 & 7 & 39 \\ \hline
\multirow{2}{*}{Llava-v1.6-34b} & Design2Code   & 0 & 46 & 3 & 54 & 0 & 33 & 139 \\
 & WebCode2M   & 0 & 5 & 1 & 12 & 0 & 15 & 33 \\
\hline
\end{tabular}}
\end{table}

\begin{tcolorbox}[colback=gray!20, colframe=gray!20, width=\columnwidth, left=0.05in, right=0.05in, top=0.05in, bottom=0.05in]
\textbf{Observation 2:} MLLMs frequently produce redundant HTML/CSS structures, and textual content, which not only increases computational overhead but also trap models in cyclical generation patterns, even leading to invalid html structures.
\end{tcolorbox}

\section{Methodology}



\begin{figure}[t]
\centering
\includegraphics[width = .95\textwidth]{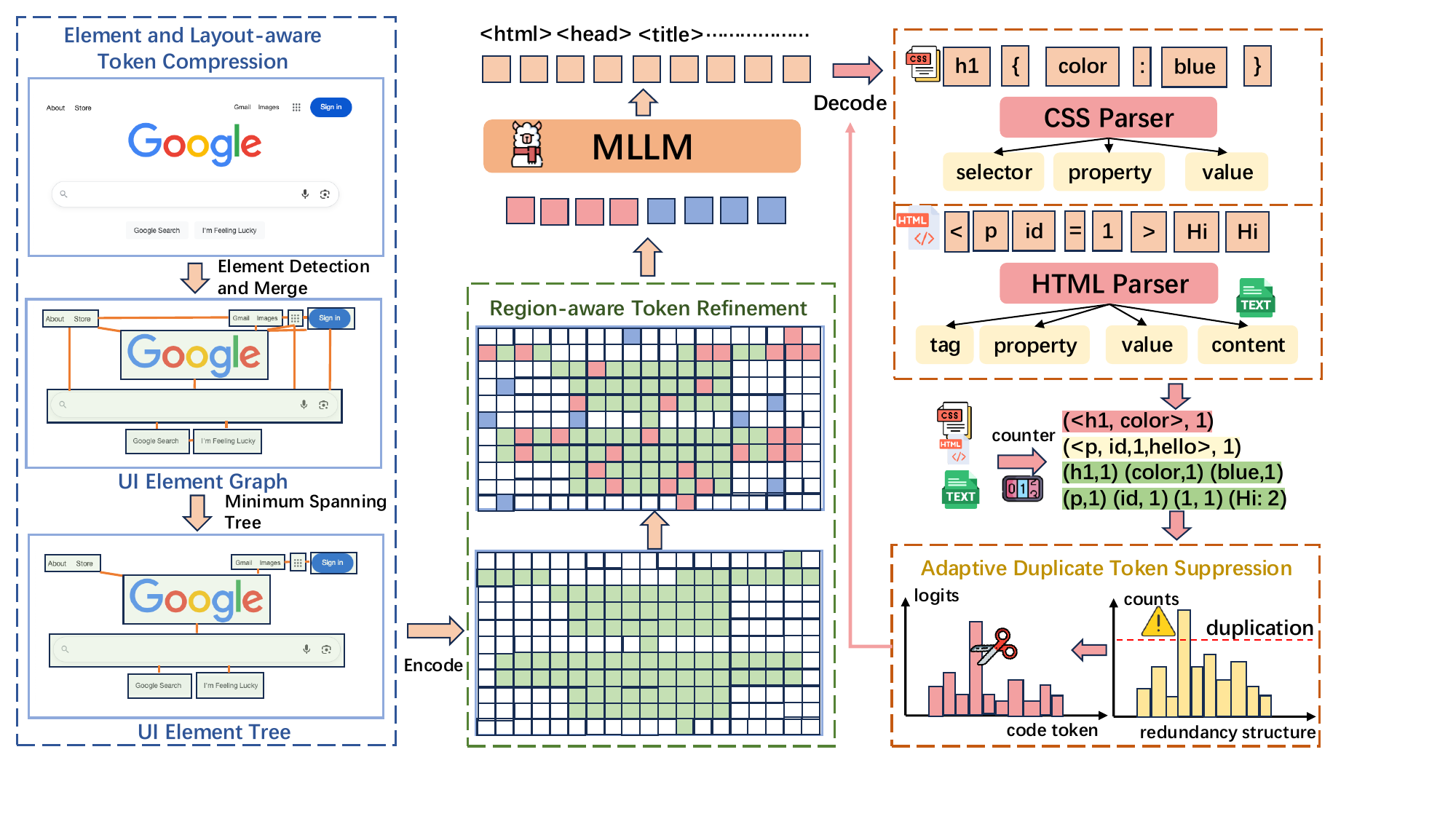}
\caption{\scheme \ framework.}
\label{fig:framework}
\end{figure}

\textbf{Overview.} Based on the two observations, we propose \scheme \ (as shown in Fig.~\ref{fig:framework})  to mitigate the image and code token redundancy. \scheme \ first applies the Element and Layout-aware Token Compression module (Section~\ref{subsec:ELTC}) to get the efficient UI representation by constructing a UI element tree. Then the Region-aware Token Refinement module (Section~\ref{subsec:RTR}) refines selected tokens by leveraging attention scores to evaluate semantic importance, discarding low-attention tokens in the selected regions while integrating high-attention tokens from unselected regions. Finally, the Adaptive Duplicate Token Suppression strategy (Section~\ref{subsec:ADTS}) first identifies the code redundancy by tracking the HTML/CSS code structure and text frequencies, then dynamically penalizes token probabilities during decoding.

\subsection{Element and Layout-Aware Token Compression}
\label{subsec:ELTC}

We aim to represent UI designs with minimal pixel while preserving semantic accuracy. This objective encompasses two critical requirements: ensuring no UI elements are omitted and maintaining correct spatial relationships between different UI components. To achieve this goal, we propose an Element and Layout-aware Token Compression method that constructs efficient UI representations by building a UI element tree structure that explicitly considers the underlying UI layout hierarchy.



\subsubsection{UI Element Tree Definition.} We first define the concept of a \textit{UI Element Graph} $G = (V, E)$ as a concise representation of user interfaces, where each node $v_i \in V$ corresponds to a distinct UI element region within the interface, including textual content, images, and components (e.g., button, input box). For any two nodes $v_i$ and $v_j$ with bounding boxes $B_i$ and $B_j$ respectively, the edge $e_{ij} \in E$ is the shortest link between two bounding boxes, the weight $w_{ij}$ represents the spatial relationship defined by the minimum distance between the boundaries of the two bounding boxes:
\begin{equation}
w_{ij} = \min_{p \in \partial B_i, q \in \partial B_j} \|p - q\|_2,
\end{equation}
where $\partial B_i$ and $\partial B_j$ denote the boundaries of bounding boxes $B_i$ and $B_j$, and $\|\cdot\|_2$ represents the Euclidean norm, $p$ and $q$ are the points on the $\partial B_i$ and $\partial B_j$. Upon this, \textit{UI Element Tree} $T = (V, E')$ is defined as a representation that consumes the least token, where $E' \subset E$, $|E'| = |V| - 1$ and $T$ is acyclic and connected. The problem is formulated as:


\begin{equation}
T^* = \arg\min_{T = (V, E')} \sum_{(v_i, v_j) \in E'} w_{ij}.
\end{equation}

\subsubsection{UI Element Tree Construction.} 
\label{subsubsec:uitree}

First, \scheme \ employs the UI Element Detection (UIED) algorithm~\cite{xie2020uied} to extract bounding boxes for all elements (i.e., text, image and component) from the UI image. Due to inter-word spacing, UIED fragments text elements into numerous small segments, a continuous sentence may be split into multiple parts. To merge these fragments, we establish a distance threshold for each text element: if the distances between text bounding boxes fall below this threshold, the fragments are merged. \revision{In the experiments, the threshold is set to 50 pixels to achieve better merging performance.} Furthermore, to ensure non-overlapping bounding boxes, we first classify the overlap relationship between boxes. In cases of inclusion, we retain the largest bounding box; for intersection relationships, we compute the minimum enclosing rectangle.


Second, \scheme \ calculates the shortest link $e_{ij}$ between two element regions (i.e., bounding boxes) $v_i$ and $v_j$, and computes the corresponding weight $w_{ij}$ to construct the UI element Graph $G$. The shortest distance computation follows three cases: (1) If two bounding boxes are horizontally separated but vertically overlapping, the shortest line segment is horizontal; (2) If two bounding boxes are vertically separated but horizontally overlapping, the shortest line segment is vertical; (3) If two bounding boxes are separated in both dimensions, the shortest line segment connects the two closest corner points.

Finally, after the UI elememt graph $G$ is construted, \scheme \ employs the Minimum Spanning Tree (MST) algorithm Kruskal~\cite{li2017kruskal}
to construct a UI element tree $T$.


\subsection{Region-Aware Token Refinement} 
\label{subsec:RTR}

Although substantial token compression is achieved in the first step, we observe that redundancy persists within element regions detected by UIED~\cite{xie2020uied}. For instance, in Fig~\ref{fig:framework}, the search input box component still contain redundant visual information, i.e., the tokens in the blank part of the search box. Similarly, some unselected background areas also contain some key information (e.g., background color), we need to add these important tokens. Therefore, we drop some unimportant tokens in the selected area and select some important tokens in the unselected area. This approach achieves substantial token compression while preserving the most semantically important visual information across both selected and unselected regions.

We assess token importance through attention score within the visual encoder. We compute attention scores as $A_{avg} = \frac{1}{H} \sum_{h=1}^{H} \text{Softmax}\left(\frac{Q_h K_h^{\top}}{\sqrt{D_h}}\right),$ where $H$ is the attention head counts, $D_h$ denotes the head dimension, and $Q_h$ and $K_h$ correspond to query and key matrices, respectively. Suppose the token number is $N$, by averaging across all attention heads, we obtain an aggregated attention matrix $A_{avg} \in \mathbb{R}^{\text{N} \times \text{N}}$, which captures the attention relationships between all token pairs. 



For CLIP~\cite{radford2021learning} visual encoder, the CLS token aggregates information from the entire image, so we use the CLS token’s attention scores to assess token importance. The importance score of token $i$ is $I_i = A_{avg}[\text{CLS}, i], \quad i = 1, 2, \ldots, N,$ where $I_i$ represents the attention weight from the CLS token to token $i$. Given the token sets $S$ (selected tokens) and $U$ (unselected tokens) from the first stage, we remove a proportion $r$ of the least important tokens from set $S$ and simultaneously select a proportion $r$ of the most important tokens from set $U$.

\subsection{Adaptive Duplicate Token Suppression}
\label{subsec:ADTS}


As shown in Section~\ref{subsec:redundant_code}, there are three types of output redundancy: css code redundancy, html code redundancy and text content redundancy. These redundancies manifest as continuous patterns in the output sequence, so we can monitor repetition frequency during the decoding phase and subsequently suppress duplications by reducing the probability of repeated tokens. \revision{Although repetition suppression is implemented as a standard technique in popular inference libraries (e.g., Huggingface Transformers~\cite{wolf2020huggingfacestransformers}), it cannot be applied in our setting. The standard repetition penalty applies a uniform penalty to all previously generated tokens, which is unsuitable for UI2Code scenarios. In particular, it will excessively suppress tokens that are expected to appear repeatedly, such as `\texttt{div}' and `\texttt{p}' tags, which naturally occur multiple times in webpages.} Accordingly, we propose an Adaptive Duplicate Token Suppression method, which consists of a repetition frequency tracking process for recording the repetition counts and an exponential penalty strategy to dynamic adjust the probabilities of the repeated token.

\begin{algorithm}[h]
\caption{Repetition Frequency Tracking}
\footnotesize
\label{algo:track}
\begin{algorithmic}[1]
\STATE \textbf{Initialize:} $css\_counts \leftarrow \{\}$; $html\_counts \leftarrow \{\}$; $text\_counts \leftarrow \{\}$
\STATE $parser\_state \leftarrow INITIAL$; $\lambda \leftarrow$ exponential decay parameter

\WHILE{MLLM generating next token}
    \STATE token $\leftarrow$ get\_next\_token()
    \STATE $sequence+=token$
    \IF{$sequence[-7:]$ = ``<style>''}
        \STATE $parser\_state \leftarrow CSS$
    \ELSIF{$sequence[-6:]$ = ``<body>''}
        \STATE $parser\_state \leftarrow HTML$
    \ENDIF
    \IF{$parser\_state = CSS$}
        \STATE CSSParser($token$, $css\_counts$, $text\_counts$)
    \ELSIF{$parser\_state = HTML$}
        \STATE HTMLParser($token$, $html\_counts$, $text\_counts$)
    \ENDIF
    \STATE Exponential\_Penalty\_Strategy($token$, $css\_counts$, $html\_counts$, $text\_counts$, $\lambda$)
\ENDWHILE

\end{algorithmic}
\end{algorithm}

\begin{algorithm}[h]
\caption{CSS Parser}
\footnotesize
\label{algo:css}
\begin{algorithmic}[1]
\Function{CSS\_PARSER}{$token$, $css\_counts$, $text\_counts$}
    \STATE \textbf{static} $cur\_selector \leftarrow$``''; $cur\_property \leftarrow$``''; $cur\_value \leftarrow$``''; $parse\_state \leftarrow SELECTOR$
    
    \IF{$token$ = ``\{''}
        \STATE $css\_counts[selector][cur\_selector] += 1$
        \STATE $parse\_state \leftarrow PROPERTY$
    \ELSIF{$token$ = ``:''}
        \STATE $css\_counts[property][cur\_property] += 1$
        \STATE $css\_counts[selector\_property][(cur\_selector, cur\_property)] += 1$
        \STATE $parse\_state \leftarrow VALUE$
    \ELSIF{$token$ = ``;'' \text{ OR } $token$ = ``\}''}
        \STATE $css\_counts[value][cur\_value] += 1$
        \IF{$token$ = ``\}''} \STATE $parse\_state \leftarrow SELECTOR$
        \ELSE \STATE $parse\_state \leftarrow PROPERTY$
        \ENDIF
    \ELSE
        \IF{$parse\_state = SELECTOR$}
            \STATE $cur\_selector += token$
            \STATE $String\_Repeat\_Detection(cur\_selector, text\_counts)$
        \ELSIF{$parse\_state = PROPERTY$}
            \STATE $cur\_property += token$
            \STATE $String\_Repeat\_Detection(cur\_property, text\_counts)$
        \ELSIF{$parse\_state = VALUE$}
            \STATE $cur\_value += token$
            \STATE $String\_Repeat\_Detection(cur\_value, text\_counts)$
        \ENDIF
    \ENDIF
\end{algorithmic}
\end{algorithm}

\begin{algorithm}[h]
\caption{HTML Parser}
\footnotesize
\label{algo:html}
\begin{algorithmic}[1]

\Function{HTML\_PARSER}{$token$, $html\_counts$, $text\_counts$}

    \STATE \textbf{static} $cur\_tag\_property\_value \leftarrow$``''; $cur\_content \leftarrow$``'';
    \STATE \textbf{static} $parse\_state \leftarrow$``''
    
    \IF{$token$ = ``<''}
        \STATE $parse\_state \leftarrow TAG\_PROPERTY\_VALUE$
        \IF{$cur\_tag\_property\_value$ != ``''}
            \STATE $html\_counts[quadruple][(cur\_tag\_property\_value, cur\_content)] += 1$
        \ENDIF
    \ELSIF{$token$ = ``>''}
        \STATE $parse\_state \leftarrow CONTENT$
    \ELSE
        \IF{$parse\_state = TAG\_PROPERTY\_VALUE$}
            \STATE $cur\_tag\_property\_value \leftarrow cur\_tag\_property\_value + token$
        \ELSIF{$parse\_state = CONTENT$}
            \STATE $cur\_content \leftarrow cur\_content + token$
            \STATE String\_Repeat\_Detection($token$, $cur\_content$, $text\_counts$)
        \ENDIF
    \ENDIF
\end{algorithmic}
\end{algorithm}

\subsubsection{Repetition Frequency Tracking}

As shown in Algorithm~\ref{algo:track}, we implement separate frequency tracking for CSS, HTML, and text content, leveraging the distinct syntactic markers in web markup languages. \revision{Because length(``<style>'')=7 and length(``<body>'')=6, during frequency tracking, if last 7 characters of the generated sequence (seq[-7:]) are ``<style>'' (line 7), we begin the css code tracking and if last 6 character (seq[-6:]) is ``<body>'' (line 8), we begin the html code tracking.}


For CSS code tracking, we track selector-property tuples since \texttt{<selector,property>} pairs uniquely define element styling specifications. Repeated \texttt{<selector,property>} represent redundant declarations, making them ideal candidates for repetition penalty. We also perform duplicate token detection on individual selector, property, and value strings, because there may be duplication of text in selector, property and value string. Algorithm~\ref{algo:css} illustrates this process: Lines 3-17 demonstrate the token-by-token parsing logic where each input token undergoes type classification to determine the appropriate parsing context. Upon encountering an opening brace (\texttt{\{}), the parser increments the frequency counter for the current selector and transitions to the \texttt{PROPERTY} state. A colon token (\texttt{:}) triggers simultaneous incrementation of the property count and the combined \texttt{<selector,property>} tuple count, followed by a state transition to \texttt{VALUE}. The parser handles statement termination through semicolon (\texttt{;}) and closing brace (\texttt{\}}) tokens, which increment the current value's frequencies and initiate transitions to either \texttt{SELECTOR} or \texttt{PROPERTY} states, respectively. For HTML code tracking, we track the quadruples \texttt{<tag,property,value,content>}, because we observe that this structure is repeated in the code. As shown in Algorithm~\ref{algo:html}, the parser maintains static variables for current tag-property-value combinations ($cur\_tag\_property\_value$) and content accumulation ($cur\_content$), along with a parsing state indicator ($parse\_state$). Upon encountering opening angle brackets (``\texttt{<}''), the parser transitions to \texttt{TAG\_PROPERTY\_VALUE} state and conditionally updates the frequency count for HTML quadruples if a valid tag-property-value combination exists. Closing angle brackets (``\texttt{>}'') trigger transitions to \texttt{CONTENT} state, enabling content parsing within HTML elements. The $String\_Repeat\_Detection$ function apply the repeated substring detection algorithm~\cite{yamamoto2001using} to record the number of times repeated substrings appear.

\subsubsection{Exponential Penalty Strategy} The decoding process for LLM begins with the computation of logits of the token. Given the hidden state $\mathbf{h}_t \in \mathbb{R}^{d}$ at position $t$, the model computes unnormalized scores (logits) for each token in the vocabulary through a linear transformation: $\mathbf{z}_t = \mathbf{W} \mathbf{h}_t + \mathbf{b}$, where $\mathbf{z}_t \in \mathbb{R}^{|V|}$ represents the logits for all tokens in vocabulary $V$, $\mathbf{W} \in \mathbb{R}^{|V| \times d}$ is the output projection matrix, and $\mathbf{b} \in \mathbb{R}^{|V|}$ is the bias vector. These logits are subsequently transformed into a probability distribution over the vocabulary using the softmax function: $P(w_t = v_i | \mathbf{x}_{<t}) = \frac{\exp(z_{t,i})}{\sum_{j=1}^{|V|} \exp(z_{t,j})},$ where $w_t$ denotes the token at position $t$, $v_i$ represents the $i$-th token in vocabulary $V$, $z_{t,i}$ is the corresponding logit, and $\mathbf{x}_{<t}$ encompasses all preceding tokens in the sequence. Finally, LLM will select the token with the highest probability to output. 
The exponential penalty strategy applies a decay mechanism that becomes more aggressive as repetition frequency increases. When duplicates occur, we will impose the following penalties on the subsequent $s$ tokens' logits to suppress repeated token:





\begin{equation}
\tilde{z}_{i} = z_{i} \cdot \lambda^{c}, \quad \forall i \in \{1, 2, \ldots, s\},
\end{equation}
where $c$ is the number of repetitions, $\lambda < 1$ is the decay factor. \revision{We modify the MLLM decoding by monitoring token frequencies. Once repetitions exceed a suppression step threshold $s$, we penalize the logits of the repeated token by multiplying them by the decay factor $\lambda$ before probability conversion, and maintain this penalty for the subsequent $s$ steps.}

\section{Experiment Setup}


\subsection{Models}

Since we need to compress the input token after the visual encoder and adjust the output decoding process, we select the popular open-source MLLMs llava-v1.6 (7B, 34B) ~\cite{liu2024llavanext} for the experiment. To improve the reproducibility of experimental results, we set the temperature of all models to 0 and fixed the random seeds to 2026. All the experiments are conducted on a Linux server equipped with 4 NVIDIA A800 80GB GPUs. We set the maximum output token length to 4096 and use the direct prompt in Design2Code~\cite{si2025design2code} for evaluation. 





\subsection{Evaluation Datasets}

We use two widely used UI2Code datasets: (1) \textbf{Design2Code~\cite{si2025design2code}}: Design2Code is a pioneering benchmark dataset comprising 484 real-world webpage screenshots paired with their corresponding front-end code (HTML/CSS), created to rigorously evaluate multimodal large language models’ ability to generate visually accurate implementations of design layouts. (2) \textbf{WebCode2M~\cite{gui2025webcode2m}}: WebCode2M is a high-quality, large-scale and real-word datasets for training and evaluating the automated webpage code generation task. We randomly select 100 webpages from WebCode2M-short, WebCode2M-mid and WebCode2M-long as the testset.

\subsection{Evaluation Metrics}

\subsubsection{Performance Metrics}


\textbf{High-level Similarity.}  
High-level metrics assess the overall fidelity of webpage appearance and code structure. (1) \textbf{CLIP Score}~\cite{radford2021learning}: visual similarity between the reference image ($I_R$) and the generated image ($I_G$) is measured by CLIP embedding similarity, denoted as $\mathbf{CLIP}(I_R, I_G)$. Features are extracted using CLIP-ViT-B/32. (2) \textbf{BLEU}~\cite{papineni2002bleu}: code similarity is evaluated by calculating the precision of n-grams in the generated HTML code with respect to the reference code, weighted by a brevity penalty to account for length differences. \revision{Here, we don't apply CodeBERT~\cite{codebert} with cosine similarity to calculate the code similarity, because: (1) it is trained on six programming languages (i.e., Go, Java, JavaScript, PHP, Python, and Ruby) and does not include HTML/CSS; and (2) its maximum input length is limited to 512 tokens, which is insufficient for UI code, as generated webpages typically require thousands of tokens.}


\textbf{Low-level Similarity.}  
High-level measures alone are insufficient to capture fine-grained discrepancies. To provide a more detailed evaluation, we adopt element-matching metrics introduced by Si et al.~\cite{si2025design2code}, which assess similarity across text, position, and color. Given reference and generated webpage screenshots, visual element blocks are detected and aligned using the Jonker-Volgenant assignment algorithm. The quality of the matches is then quantified by: (1) \textbf{Block-match}: the ratio of matched block areas to total block areas, penalizing missing or hallucinated elements; (2) \textbf{Text similarity}: character overlap measured by the Sørensen-Dice coefficient; (3) \textbf{Color similarity}: perceptual differences computed using the CIEDE2000 formula; (4) \textbf{Position similarity}: alignment accuracy based on block center locations.

\subsubsection{Efficiency Metrics}

We evaluate the efficiency of MLLMs on the UI2Code task using the following metrics, which capture both computational complexity and temporal performance: (1) \textbf{Compression Ratio}: $R=\frac{C_{compressed}}{C_{original}}$, where $C_{compressed}$ and $C_{original}$ denote the number of visual tokens in the compressed and original UI image respectively. For our method, since the proportion of tokens compressed on each UI image is different, we report the average compression rate of the entire dataset. (2) \textbf{Floating Point Operations (FLOPs):} this metric quantifies the total number of arithmetic operations performed by the model during the complete inference process. Since compression is applied subsequent to the visual encoder, we specifically calculate the FLOPs of the LLM backbone to ensure fair comparison across different architectures. (3) \textbf{Prefill Time:} This measures the latency required to generate the first output token, encompassing input preprocessing, key-value (KV) cache initialization, and the initial forward pass computation. (4) \textbf{Inference Time:} This represents the end-to-end latency from input reception to complete output sequence generation, providing a comprehensive measure of the model's real-world deployment efficiency.

\subsubsection{\revision{Human Evaluation Metrics}}

\revision{While automatic metrics provide a fine-grained assessment of model performance, evaluating user perceptions of the generated webpages is also essential. Following Design2Code~\cite{si2025design2code}, we conduct a human evaluation, as detailed below.} \revision{(1) \textbf{Participant Information and Recruitment}. We recruit five annotators through a local university mailing list. All annotators hold at least a B.S. degree in computer science or software engineering and have more than two years of web-development experience. (2) \textbf{Evaluation Procedure}. Step 1. We randomly sample 50 pages from Design2Code and 50 from WebCode2M. Each annotator evaluate 100 unique pairs (Vanilla vs. one compression method) in a single round. Step 2. Annotators are required to carefully read the pairwise comparison guidelines\footnote{https://github.com/WebPAI/EfficientUICoder/blob/main/assets/Human\_Evaluation.md}. Step 3. For each pair, annotators select ``Example 1 better'', ``Example 2 better'', or ``Tie'', following the detailed guidelines. Annotators are blind to which method produced each webpage, and pairs are presented in randomized order. (3) \textbf{Inter-rater Reliability}. To assess the consistency and reliability of the human annotations, we compute Krippendorff’s alpha. The resulting value is 0.8308, which exceeds the commonly adopted threshold of 0.8, indicating a high level of inter-annotator agreement.}

\subsection{Baselines}

\begin{itemize}[leftmargin=*]
    \item \textbf{Vanilla} denotes the original model without any token compression.
    \item \textbf{Random} means randomly selecting a certain proportion of tokens from the UI image.
    \item \textbf{FastV}~\cite{chen2024fastv} prunes low-attention visual tokens after a designated LLM layer.
    \item \textbf{Pdrop}~\cite{xing2024pyramiddrop} partition the MLLMs into several stages and drop part of the image tokens at the end of each stage with a pre-defined ratio. The dropping is based on the token similarity calculation. 
    \item \textbf{VisionZip}~\cite{yang2025visionzip} selects dominant tokens by visual encoder's attention score, discarding a certain proportion of visual tokens before inputting the MLLMs.
\end{itemize}


\subsection{Research Questions}


    



\begin{itemize}[leftmargin=*]
    \item (RQ1) \textbf{Performance}. Can \scheme \ effectively compress token while preserving the UI2Code performance?
    \item (RQ2) \textbf{Efficiency}. What is the efficiency benefit of \scheme \ in UI2Code tasks?
    \item (RQ3) \textbf{Ablation study}. How does different parts contribute to \scheme's performance?    
    \item (RQ4) \textbf{Parameter study}. How do key parameters affect the performance of \scheme?
    \item (RQ5) \textbf{Case Study.} Why does \scheme \ work?
\end{itemize}

\section{Experiment Results}


\subsection{Performance Comparison (RQ1)}
\label{subsec:RQ1}

Table~\ref{tab:RQ1} presents the performance comparison of different compression methods. For \scheme, after applying the ELTC and RTR modules, we achieve compression ratios of \textcolor{green!50!black}{60.36\%} on the Design2Code dataset and \textcolor{green!50!black}{55.86\%} on the Webcode2M dataset. To ensure fair comparison, we configure all baseline methods to achieve the same compression ratios. We can make the following observations: (1) \textbf{\scheme \ demonstrates superior performance while compressing 55\%-60\% of tokens.} Our method consistently outperforms existing compression techniques across most evaluation metrics. This superiority stems from our structured approach to token selection: while baseline methods either select tokens randomly or rely solely on attention mechanisms without considering UI structure, EfficientUICoder's ELTC module performs selection based on UI elements and layout hierarchy, while the RTR module incorporates semantic importance. Remarkably, our method even surpasses the vanilla version in several metrics. We attribute this improvement to two factors: first, EfficientUICoder's token selection mechanism enables the model to focus more effectively on critical UI elements and layout information; second, the ADTS component efficiently reduces duplicate tokens, facilitating the generation of more coherent and valid webpages. (2) \textbf{\scheme \ significantly reduces redundant code generation.} The RS metric demonstrates substantial improvements: for the 7b model, redundant samples decrease from 148 to 64 on Design2Code (56.8\% reduction) and from 39 to 15 on Webcode2M (61.5\% reduction). These results validate that the ADTS module effectively suppresses redundant code generation, leading to more concise and efficient outputs. \revision{(3) EfficientUICoder consistently outperforms baselines across datasets and model sizes. Compared with vanilla, performance differences are insignificant when vanilla averages are higher, while several gains are statistically significant when EfficientUICoder leads, indicating that it maintains or even surpasses near-uncompressed performance.}

\begin{table*}[ht]
\centering
\setlength{\tabcolsep}{0.12em}
\caption{Performance Comparison of different Compression Methods on Two Datasets. On the Design2Code dataset, the compression ratio is \textcolor{green!50!black}{60.36\%}. On the Webcode2M dataset the compression ratio is \textcolor{green!50!black}{55.86\%}. Bold values indicate the optimal performance, and underlined values indicate the second best performance. RS is the number of samples with redundant codes. \revision{Rows marked with $p$-value show statistical significance of Mann-Whitney U test comparing EfficientUICoder vs. each baseline ($p < 0.05$ indicates statistical significance). EUC denotes EfficientUICoder.}}
\label{tab:RQ1}
\resizebox{0.99\textwidth}{!}{
\begin{tabular}{l|ccccccc|ccccccc}
\hline
\multirow{2}{*}{Method} & \multicolumn{7}{c|}{\textbf{Design2Code}} & \multicolumn{7}{c}{\textbf{Webcode2M}} \\
\cline{2-15}
& Block & Text & Position & Color & CLIP & BLEU & RS & Block & Text & Position & Color & CLIP & BLEU & RS \\
\hline
\multicolumn{15}{c}{\it Llava-v1.6-7b} \\ \hline
Vanilla          & \textbf{0.3675} & \underline{0.7275} & \underline{0.5655} & \underline{0.5124} & \underline{0.6916} & \underline{0.1667} & 148 & \textbf{0.2843} & \underline{0.7408} & \underline{0.5728} & \textbf{0.5056} & \underline{0.7123} & \underline{0.1190} & 39 \\
\revision{$p$-value} & \revision{2.6e-1} & \revision{5.5e-2} & \revision{4.7e-2} & \revision{8.6e-1} & \revision{2.3e-2} & \revision{2.0e-2} & \revision{-} & \revision{7.4e-1} & \revision{7.3e-1} & \revision{2.1e-1} & \revision{7.3e-1} & \revision{1.1e-1} & \revision{1.4e-1} & \revision{-} \\ \hline
Random           & 0.1699 & 0.5998 & 0.5001 & 0.4599 & 0.6492 & 0.0413 & 144 & 0.1449 & 0.5854 & 0.4829 & 0.4418 & 0.6373 & 0.0301 & \underline{34} \\
\revision{$p$-value} & \revision{2.2e-17} & \revision{7.4e-22} & \revision{8.3e-10} & \revision{7.4e-3} & \revision{8.1e-4} & \revision{1.2e-32} & \revision{-} & \revision{1.6e-4} & \revision{7.9e-6} & \revision{1.4e-4} & \revision{2.5e-1} & \revision{1.5e-2} & \revision{2.5e-7} & \revision{-} \\ \hline
FastV            & 0.2972 & 0.6141 & 0.4769 & 0.4209 & 0.5886 & 0.1154 & \underline{132} & 0.2052 & 0.6785 & 0.5291 & 0.4546 & 0.6594 & 0.0832 & 40 \\ 
\revision{$p$-value} & \revision{1.2e-3} & \revision{6.3e-3} & \revision{7.0e-9} & \revision{8.0e-6} & \revision{2.2e-8} & \revision{2.1e-12} & \revision{-} & \revision{5.6e-2} & \revision{4.1e-1} & \revision{1.8e-2} & \revision{3.0e-1} & \revision{2.5e-2} & \revision{9.1e-3} & \revision{-} \\ \hline
Pdrop            & 0.1907 & 0.6342 & 0.5108 & 0.4657 & 0.6587 & 0.0602 & 141 & 0.1657 & 0.6738 & 0.5393 & 0.4628 & 0.6866 & 0.0698 & 35 \\
\revision{$p$-value} & \revision{5.5e-14} & \revision{1.5e-13} & \revision{2.0e-9} & \revision{1.5e-2} & \revision{2.6e-4} & \revision{2.6e-21} & \revision{-} & \revision{4.4e-3} & \revision{2.2e-2} & \revision{6.9e-3} & \revision{3.5e-1} & \revision{1.0e-1} & \revision{1.1e-2} & \revision{-} \\ \hline
Visionzip        & 0.3242 & 0.6933 & 0.5357 & 0.4877 & 0.6694 & 0.1388 & 204 & 0.1973 & 0.6923 & 0.5607 & 0.4625 & 0.6799 & 0.0705 & 53 \\
\revision{$p$-value} & \revision{3.0e-1} & \revision{6.9e-1} & \revision{5.5e-5} & \revision{3.7e-1} & \revision{3.5e-3} & \revision{1.4e-6} & \revision{-} & \revision{5.6e-2} & \revision{3.5e-1} & \revision{1.4e-1} & \revision{6.1e-1} & \revision{1.2e-2} & \revision{7.6e-4} & \revision{-} \\ \hline
EUC & \underline{0.3374} & \textbf{0.7333} & \textbf{0.5939} & \textbf{0.5125} & \textbf{0.7172} & \textbf{0.1855} & \textbf{64} & \underline{0.2718} & \textbf{0.7445} & \textbf{0.6170} & \underline{0.4909} & \textbf{0.7393} & \textbf{0.1338} & \textbf{15} \\ \hline
\multicolumn{15}{c}{\it Llava-v1.6-34b} \\ \hline
Vanilla          & \textbf{0.5516} & 0.7914 & 0.6292 & 0.5842 & 0.7343 & \textbf{0.2914} & 136 & \textbf{0.4688} & \textbf{0.8207} & \textbf{0.6441} & \textbf{0.5546} & \textbf{0.7554} & \textbf{0.2689} & 33 \\
\revision{$p$-value} & \revision{9.2e-2} & \revision{1.1e-3} & \revision{9.1e-1} & \revision{5.7e-1} & \revision{8.6e-2} & \revision{1.1e-1} & \revision{-} & \revision{5.4e-1} & \revision{1.9e-1} & \revision{1.2e-1} & \revision{8.9e-1} & \revision{5.1e-1} & \revision{9.1e-1} & \revision{-} \\ \hline
Random           & 0.2108 & 0.5406 & 0.4629 & 0.4297 & 0.5845 & 0.0552 & 98 & 0.1291 & 0.4769 & 0.4491 & 0.3385 & 0.5607 & 0.0443 & 30 \\
\revision{$p$-value} & \revision{1.5e-50} & \revision{1.5e-53} & \revision{2.8e-30} & \revision{6.7e-24} & \revision{8.8e-23} & \revision{6.2e-57} & \revision{-} & \revision{1.5e-12} & \revision{2.5e-15} & \revision{6.9e-6} & \revision{2.5e-8} & \revision{3.0e-8} & \revision{7.5e-14} & \revision{-} \\ \hline
FastV            & 0.3161 & 0.6913 & 0.5789 & 0.5289 & 0.7085 & 0.1025 & 125 & 0.2719 & 0.6691 & 0.5518 & 0.4620 & 0.6878 & 0.0774 & 34 \\
\revision{$p$-value} & \revision{2.8e-22} & \revision{4.1e-24} & \revision{1.1e-12} & \revision{4.3e-9} & \revision{3.5e-13} & \revision{2.4e-32} & \revision{-} & \revision{5.3e-5} & \revision{1.5e-8} & \revision{1.9e-3} & \revision{3.3e-4} & \revision{2.7e-5} & \revision{1.0e-8} & \revision{-} \\ \hline
Pdrop            & 0.4117 & 0.7071 & 0.5744 & 0.5124 & 0.6876 & 0.1577 & 111 & 0.3747 & 0.6968 & 0.5750 & 0.4644 & 0.6892 & 0.1599 & 23 \\
\revision{$p$-value} & \revision{1.0e-29} & \revision{1.7e-28} & \revision{9.3e-27} & \revision{4.3e-27} & \revision{1.0e-29} & \revision{3.0e-35} & \revision{-} & \revision{3.1e-4} & \revision{1.7e-6} & \revision{4.2e-4} & \revision{4.6e-6} & \revision{7.1e-6} & \revision{7.9e-6} & \revision{-} \\ \hline
Visionzip        & 0.4824 & \underline{0.8040} & \underline{0.6333} & \underline{0.6013} & \underline{0.7620} & 0.2313 & \underline{75} & 0.4176 & 0.7077 & \underline{0.5959} & 0.4886 & 0.6903 & 0.2117 & \underline{22} \\
\revision{$p$-value} & \revision{5.2e-8} & \revision{6.4e-5} & \revision{9.6e-13} & \revision{2.5e-8} & \revision{1.4e-15} & \revision{1.6e-9} & \revision{-} & \revision{4.9e-4} & \revision{5.7e-6} & \revision{9.1e-4} & \revision{1.9e-5} & \revision{2.2e-6} & \revision{5.3e-5} & \revision{-} \\ \hline
EUC & \underline{0.5318} & \textbf{0.8113} & \textbf{0.6643} & \textbf{0.6110} & \textbf{0.7864} & \underline{0.2547} & \textbf{16} & \underline{0.4382} & \underline{0.7533} & 0.5907 & \underline{0.5313} & \underline{0.7270} & \underline{0.2634} & \textbf{2} \\
\hline
\end{tabular}}
\end{table*}

\begin{figure}[t]
    \centering
    \begin{minipage}{0.48\linewidth}
        \centering
        \subfigure[Llava-v1.6-7b.]{
            \includegraphics[width=0.45\linewidth]{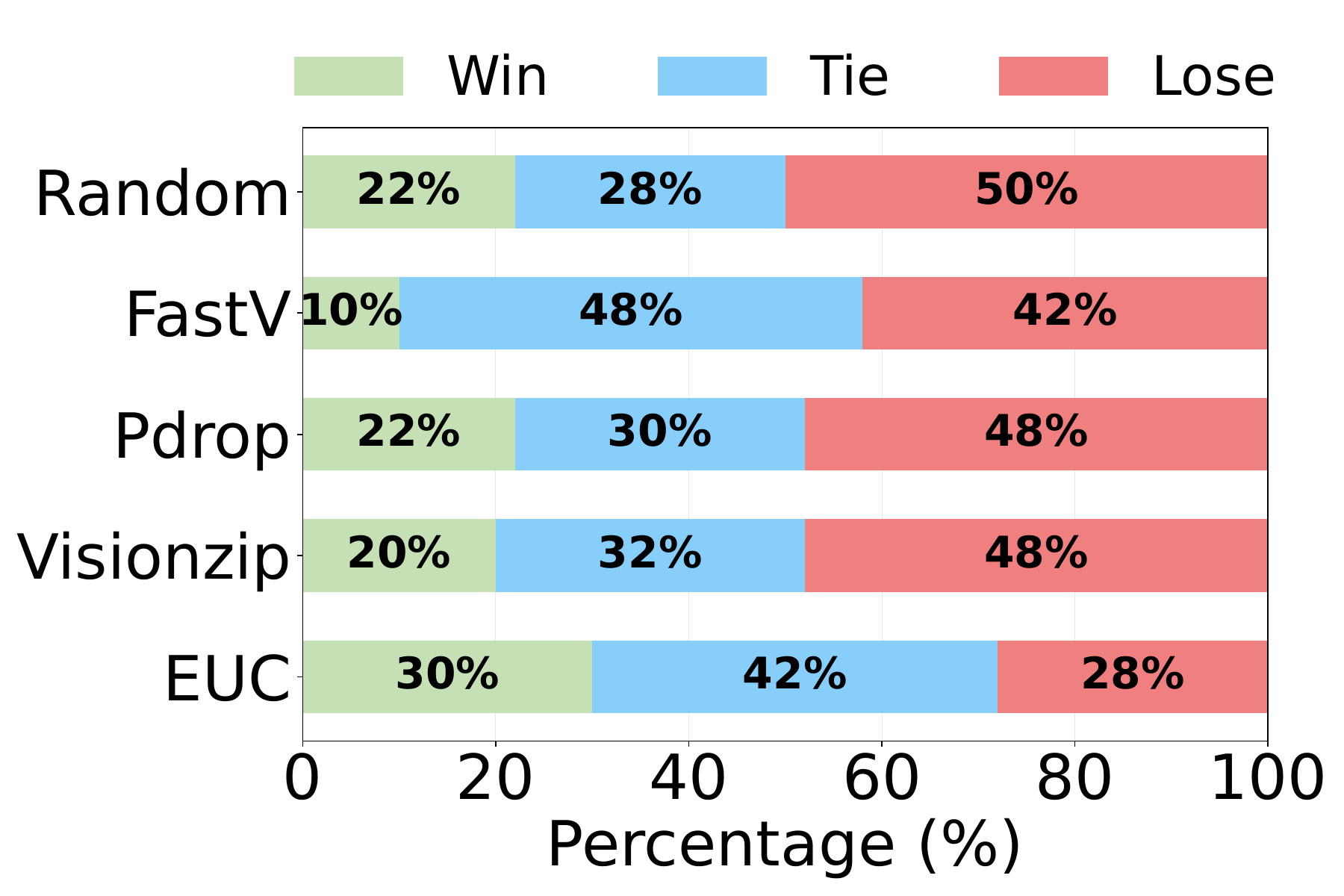}
        }
        \subfigure[Llava-v1.6-34b.]{
            \includegraphics[width=0.45\linewidth]{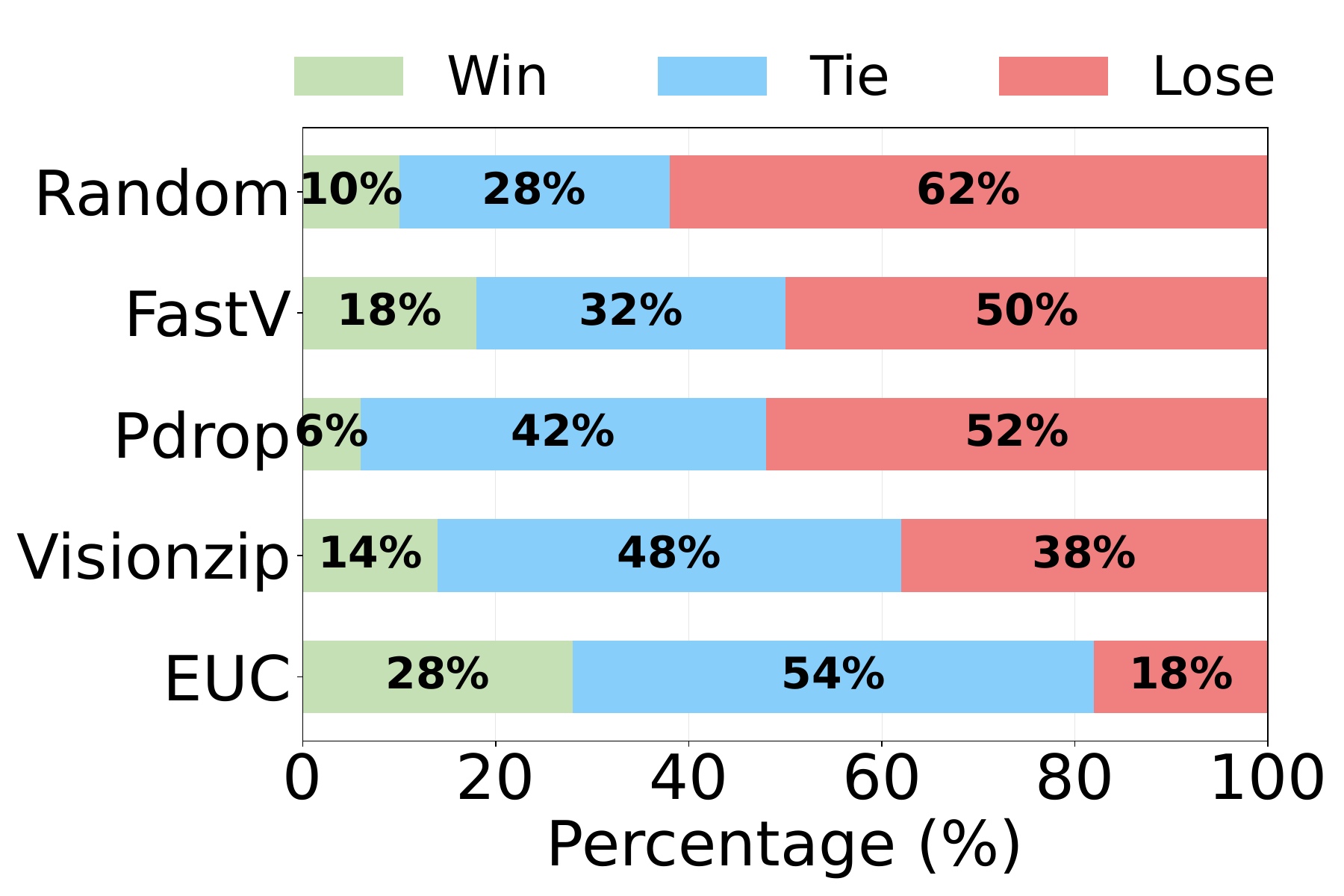}
        }
        \caption{Human Evaluation on Design2Code.}
        \label{fig:humaneval_d2c}
    \end{minipage}
    \begin{minipage}{0.48\linewidth}
        \centering
        \subfigure[Llava-v1.6-7b.]{
            \includegraphics[width=0.45\linewidth]{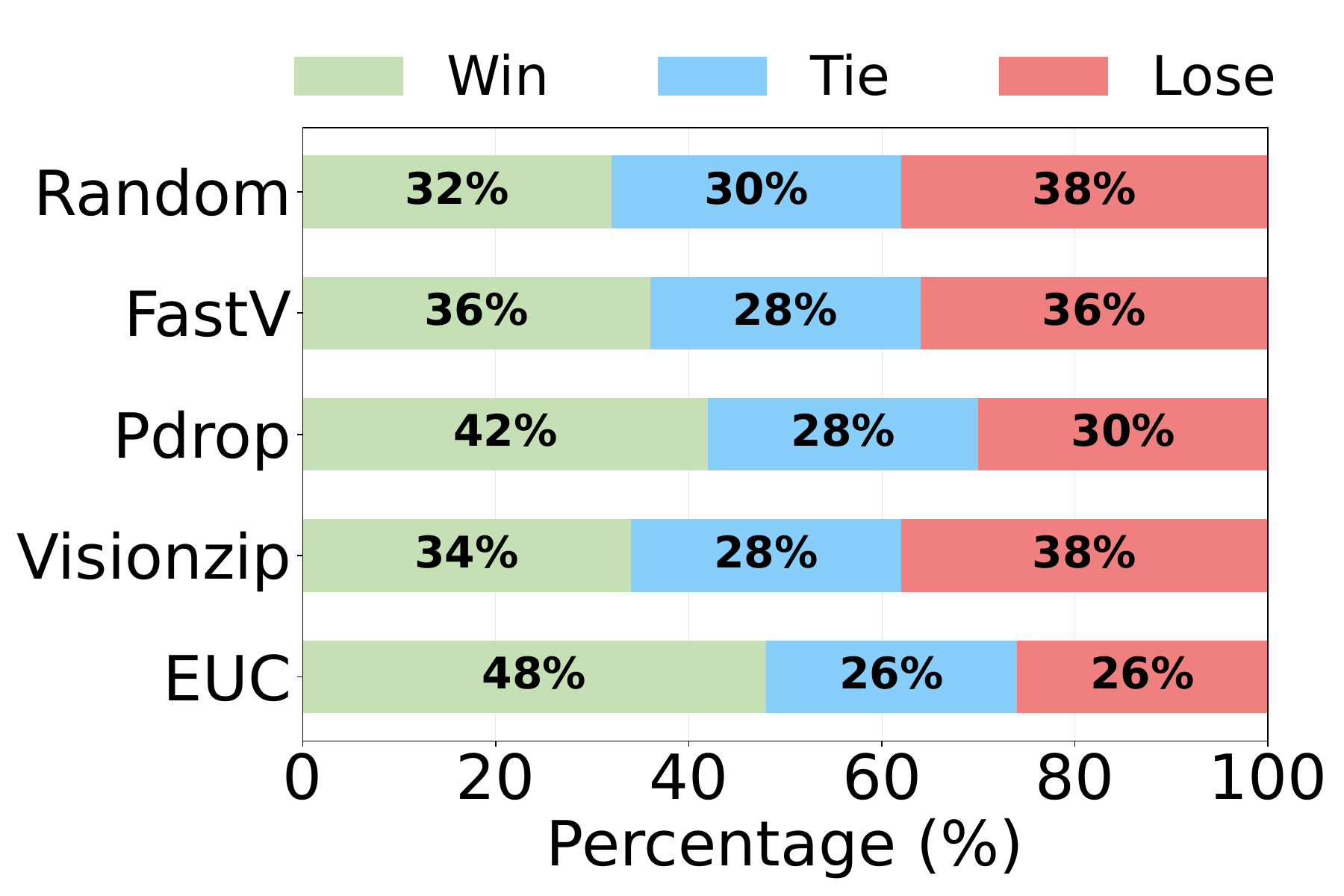}
        }
        \subfigure[Llava-v1.6-34b.]{
            \includegraphics[width=0.45\linewidth]{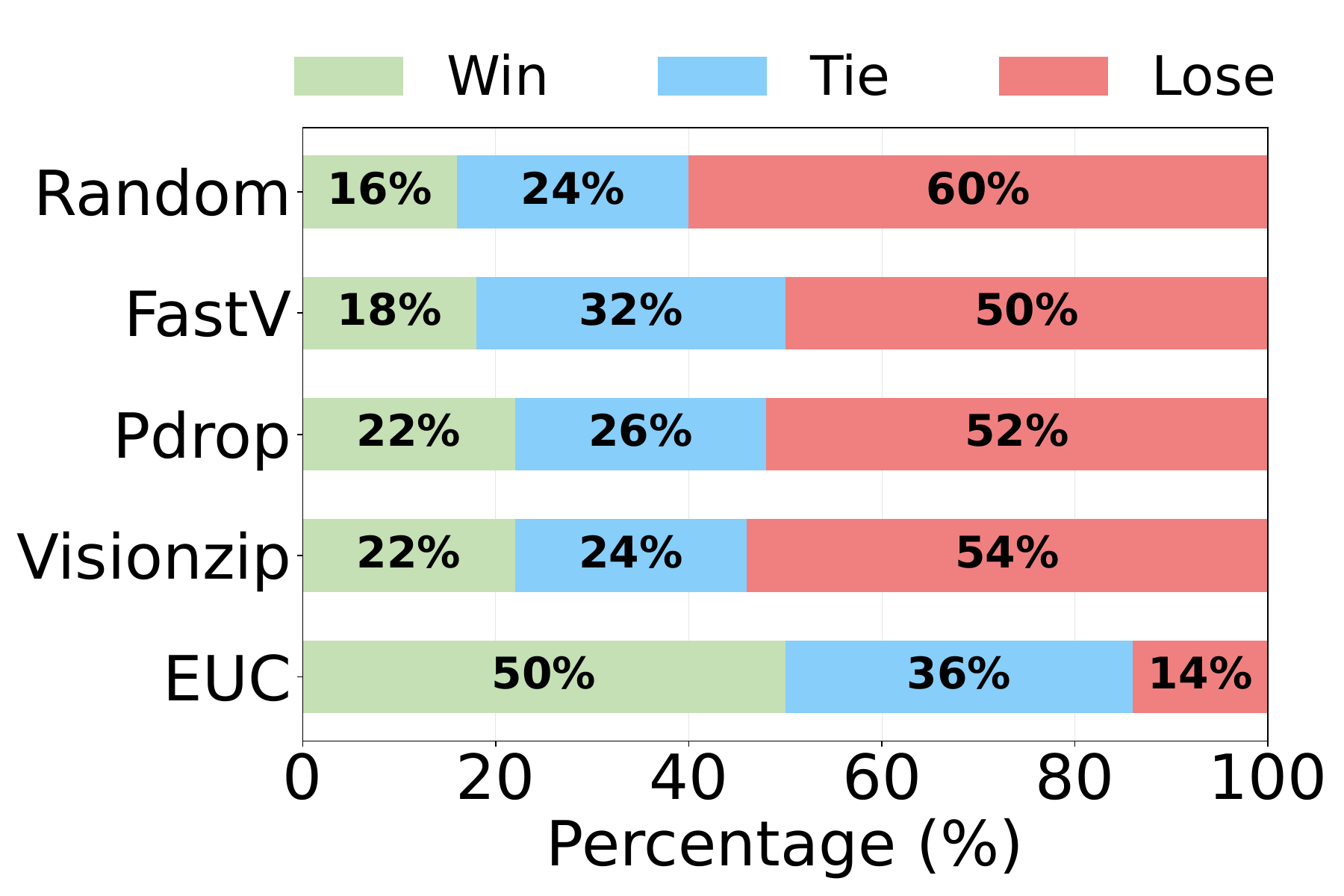}
        }
        \caption{Human Evaluation on WebCode2M.}
        \label{fig:humaneval_webcode2m}
    \end{minipage}
\end{figure}

\textbf{Human Evaluation}. 
We conduct human evaluation on 50 webpages from Design2Code and 50 webpages from WebCode2M. Fig.~\ref{fig:humaneval_d2c} and Fig.~\ref{fig:humaneval_webcode2m} (EUC denotes \scheme) present the pairwise preference evaluation results, where five annotators compare different methods against the Vanilla using majority voting. Higher win rates and lower loss rates indicate superior code quality as assessed by human evaluators. The results demonstrate that \scheme \ consistently outperforms all baseline methods across both datasets. Importantly, these human evaluation results corroborate our automatic metrics, confirming the validity of our automated metrics. \revision{As shown in Table~\ref{tab:human_significance}, the consistently low p-values indicate that human evaluators judge EfficientUICoder's outputs to be significantly higher quality than those of other compression methods.
}

\begin{table}[t]
\centering
\caption{\revision{Human evaluation’s Mann-Whitney-U test results between EUC and baselines. EUC denotes \scheme, "Y" denotes significant (p-value<0.05) and "N" denotes not significant (p-value>=0.05).}}
\label{tab:human_significance}
\resizebox{.9\textwidth}{!}{
\begin{tabular}{l|cccc}
\hline
\textbf{\revision{Dataset}} & \textbf{\revision{EUC vs. random}} & \textbf{\revision{EUC vs. fastv}} & \textbf{\revision{EUC vs. pdrop}} & \textbf{\revision{EUC vs. visionzip}} \\
\hline
\revision{Design2Code} & \revision{1.88e-05 (Y)} & \revision{3.22e-05 (Y)} & \revision{8.27e-05 (Y)} & \revision{5.60e-05 (Y)} \\
\revision{WebCode2M}   & \revision{1.63e-05 (Y)} & \revision{3.93e-03 (Y)} & \revision{3.82e-05 (Y)} & \revision{1.49e-02 (Y)} \\
\hline
\end{tabular}}
\end{table}

\begin{tcolorbox}[colback=gray!20, colframe=gray!20, width=\columnwidth, left=0.05in, right=0.05in, top=0.05in, bottom=0.05in]
\textbf{Answer to RQ1:} 
\scheme \ effectively achieves 55\%-60\% image token compression and 56\%-94\% code redundancy reduction while maintaining or improving generation performance.
\end{tcolorbox}


\subsection{Efficiency Comparison (RQ2)}

Table~\ref{tab:efficiency} presents the efficiency analysis of different compression methods on Llava-v1.6-34b across both datasets. Several key observations emerge from this analysis.(1) \textbf{\scheme \ achieves superior computational savings.} Our method demonstrates the highest FLOPs reduction of 44.9\% on Design2Code and 42.8\% on Webcode2M, significantly outperforming other techniques. This translates to substantial reductions in computational requirements and energy consumption. (2) \textbf{\scheme \ achieves significant reduction in output tokens.} EfficientUICoder dramatically reduces generated tokens while maintaining quality: from 2041 to 1195 tokens on Design2Code (41.4\% reduction) and from 2116 to 1340 on Webcode2M (36.7\% reduction). This indicates effective elimination of redundant code patterns. (3) \textbf{Superior inference and prefill acceleration.} Our method achieves the most significant inference time improvements: 48.8\% reduction on Design2Code and 45.1\% on Webcode2M, substantially exceeding other methods. Additionally, EfficientUICoder demonstrates competitive prefill time performance with 46.6\% and 45.9\% improvements respectively. This dual acceleration in both prefill and inference phases is crucial for real-world deployment scenarios. \revision{\scheme \ also shows statistically significant superiority across both datasets when compared to compression-based baselines and vanilla. It denotes that \scheme \ demonstrates significant efficiency improvements.}

\begin{table*}[t] 
\centering 
\setlength{\tabcolsep}{0.12em} 
\caption{Efficiency comparison on \textit{Llava-v1.6-34b} for two datasets. GT denotes the number of generated tokens, PT denotes the prefill time, and IT denotes the inference time. \revision{P-value of Mann-Whitney-U test between \scheme \ and baselines are reported in the row below each baselines. EUC denotes \scheme, "Y" denotes significant (p-value<0.05) and "N" denotes not significant (p-value>=0.05).}} 
\label{tab:efficiency}
\resizebox{\textwidth}{!}{
\begin{tabular}{l|cccc|cccc} 
\hline 
\multirow{2}{*}{Method} & \multicolumn{4}{c|}{\textbf{Design2Code}} & \multicolumn{4}{c}{\textbf{Webcode2M}} \\ 
\cline{2-9} 
& FLOPs (T) & GT & PT (ms) & IT (s) & FLOPs (T) & GT & PT (ms) & IT (s) \\ 
\hline 

Vanilla          
& 184.1 & 2041 & 1012 & 160 
& 195.0 & 2116 & 1102 & 173  \\ 
\revision{$p$-value} 
& \revision{5.53e-129 (Y)} & \revision{1.10e-14 (Y)} & \revision{1.18e-158 (Y)} & \revision{3.58e-28 (Y)}
& \revision{1.53e-27 (Y)} & \revision{1.42e-03 (Y)} & \revision{2.89e-34 (Y)} & \revision{5.34e-06 (Y)} \\ \hline

Random          
& 121.7 \textcolor{green!50!black}{$\downarrow$ 33.9\%} & \underline{1710} \textcolor{green!50!black}{$\downarrow$ 16.2\%} 
& 549 \textcolor{green!50!black}{$\downarrow$ 45.8\%} & \underline{122} \textcolor{green!50!black}{$\downarrow$ 23.8\%} 
& 144.3 \textcolor{green!50!black}{$\downarrow$ 26.0\%} & 2144 \textcolor{red}{$\uparrow$ 1.3\%} 
& 607 \textcolor{green!50!black}{$\downarrow$ 44.9\%} & 156 \textcolor{green!50!black}{$\downarrow$ 9.8\%} \\ 

\revision{$p$-value} 
& \revision{1.48e-04 (Y)} & \revision{9.15e-09 (Y)} & \revision{5.79e-06 (Y)} & \revision{3.41e-11 (Y)}
& \revision{5.58e-02 (N)} & \revision{1.17e-02 (Y)} & \revision{2.09e-02 (Y)} & \revision{2.64e-03 (Y)} \\ \hline

FastV          
& \underline{119.9} \textcolor{green!50!black}{$\downarrow$ 34.9\%} & 1864 \textcolor{green!50!black}{$\downarrow$ 8.7\%} 
& 606 \textcolor{green!50!black}{$\downarrow$ 40.1\%} & 130 \textcolor{green!50!black}{$\downarrow$ 18.8\%} 
& 138.0 \textcolor{green!50!black}{$\downarrow$ 29.2\%} & 2136 \textcolor{red}{$\uparrow$ 0.9\%} 
& 734 \textcolor{green!50!black}{$\downarrow$ 33.4\%} & 158 \textcolor{green!50!black}{$\downarrow$ 8.7\%} \\ 

\revision{$p$-value} 
& \revision{9.43e-07 (Y)} & \revision{1.46e-02 (Y)} & \revision{6.93e-25 (Y)} & \revision{4.38e-03 (Y)}
& \revision{1.33e-02 (Y)} & \revision{7.31e-01 (N)} & \revision{3.85e-14 (Y)} & \revision{5.78e-01 (N)} \\ \hline

Pdrop          
& 122.7 \textcolor{green!50!black}{$\downarrow$ 33.4\%} & 1958 \textcolor{green!50!black}{$\downarrow$ 4.1\%} 
& \textbf{491} \textcolor{green!50!black}{$\downarrow$ 51.5\%} & 137 \textcolor{green!50!black}{$\downarrow$ 14.4\%} 
& \underline{131.2} \textcolor{green!50!black}{$\downarrow$ 32.7\%} & 1980 \textcolor{green!50!black}{$\downarrow$ 6.4\%} 
& \textbf{582} \textcolor{green!50!black}{$\downarrow$ 47.2\%} & 144 \textcolor{green!50!black}{$\downarrow$ 16.8\%} \\ 

\revision{$p$-value} 
& \revision{7.72e-18 (Y)} & \revision{7.62e-08 (Y)} & \revision{1.21e-33 (Y)} & \revision{2.71e-11 (Y)}
& \revision{3.21e-04 (Y)} & \revision{8.30e-03 (Y)} & \revision{9.16e-03 (Y)} & \revision{1.68e-03 (Y)} \\ \hline

Visionzip          
& 125.4 \textcolor{green!50!black}{$\downarrow$ 31.9\%} & 1807 \textcolor{green!50!black}{$\downarrow$ 11.5\%} 
& 544 \textcolor{green!50!black}{$\downarrow$ 46.2\%} & 127 \textcolor{green!50!black}{$\downarrow$ 20.6\%} 
& 137.2 \textcolor{green!50!black}{$\downarrow$ 29.6\%} & \underline{1960} \textcolor{green!50!black}{$\downarrow$ 7.4\%} 
& \underline{592} \textcolor{green!50!black}{$\downarrow$ 46.3\%} & \underline{142} \textcolor{green!50!black}{$\downarrow$ 17.9\%} \\ 

\revision{$p$-value} 
& \revision{2.71e-03 (Y)} & \revision{3.76e-05 (Y)} & \revision{6.53e-06 (Y)} & \revision{1.07e-07 (Y)}
& \revision{1.24e-02 (Y)} & \revision{1.02e-02 (Y)} & \revision{9.56e-03 (Y)} & \revision{1.88e-03 (Y)} \\ \hline 

EUC          
& \textbf{101.5} \textcolor{green!50!black}{$\downarrow$ 44.9\%} & \textbf{1195} \textcolor{green!50!black}{$\downarrow$ 41.4\%} 
& \underline{540} \textcolor{green!50!black}{$\downarrow$ 46.6\%} & \textbf{82} \textcolor{green!50!black}{$\downarrow$ 48.8\%} 
& \textbf{111.6} \textcolor{green!50!black}{$\downarrow$ 42.8\%} & \textbf{1340} \textcolor{green!50!black}{$\downarrow$ 36.7\%} 
& \underline{596} \textcolor{green!50!black}{$\downarrow$ 45.9\%} & \textbf{95} \textcolor{green!50!black}{$\downarrow$ 45.1\%} \\ \hline 
\end{tabular}}
\end{table*}


\begin{tcolorbox}[colback=gray!20, colframe=gray!20, width=\columnwidth, left=0.05in, right=0.05in, top=0.05in, bottom=0.05in]
\textbf{Answer to RQ2:} 
EfficientUICoder achieves superior efficiency improvements across all metrics, reducing computational cost by up to 44.9\%, generated tokens by up to 41.4\%, prefill time by up to 46.6\%, and inference time by up to 48.8\%.
\end{tcolorbox}

\subsection{Ablation Study (RQ3)}

\scheme \ comprises three core components: Element and Layout-aware Token Compression (\textbf{ELTC}), Region-aware Token Refinement (\textbf{RTR}), and Adaptive Duplicate Token Suppression (\textbf{ADTS}). To evaluate the contribution of each component, we conduct an ablation study using five variants of \scheme ($C_0$-$C_5$). {\color[HTML]{00CD66} Y} indicates the inclusion of the corresponding component, while {\color[HTML]{FF0000} X} denotes its removal. Configuration $C_5$ represents the complete \scheme \ with all three components enabled. Since RTR requires the token selection from ELTC to function properly, we replace RTR with a high-attention-score token selection strategy when ELTC is disabled to ensure fair comparison.  First, Table~\ref{tab:ablation} demonstrates that each individual component contributes positively to the overall performance of \scheme. The combination of all three components ($C_4$) achieves superior performance and efficiency compared to any partial configuration, highlighting the complementary nature of these components in optimizing UI code generation tasks. \revision{Second, $C_1$ and $C_4$ provide ablations that separately examine input compression (ELTC + RTR) and the repetition-suppression mechanism (ADTS). The results show: (1) Removing input compression substantially increases FLOPs, prefill time, and overall inference latency, while also degrading UI2Code performance. Redundant visual tokens distract the model from key UI elements, reducing accuracy, and the inflated number of image tokens further increases computational cost (FLOPs), leading to slower prefill and inference. (2) Removing ADTS decreases the performance and increases inference time, because duplicate contents affect the webpages' appearance and endless repetition severely hinders UI2Code's end-to-end inference time latency.} \revision{In summary, input compression and repetition suppression work in tandem: compression streamlines visual inputs, while ADTS ensures clean, non-repetitive outputs. Removing either component degrades both efficiency and accuracy, showing that the two mechanisms are complementary.}

\begin{table}[h]
\caption{The performance and efficiency of 5 variants ($C_0$-$C_5$) on \textit{Llava-v1.6-7b} for WebCode2M.}
\label{tab:ablation}
\setlength{\tabcolsep}{0.11em}
\resizebox{.9\textwidth}{!}{
\begin{tabular}{@{}c|cccc|cccccc|ccc@{}}
\toprule
Datasets & ELTC                     & RTR                     & ADTS                     & & Block     & Text     & Position     & Color & CLIP & BLEU &FLOPs (T) & PT (ms) & IT (s)   \\ \midrule
\multirow{5}{*}{WebCode2M} & {\color[HTML]{FF0000} X} & {\color[HTML]{FF0000} X} & {\color[HTML]{FF0000} X} & $C_0$    & \underline{0.2843} & 0.7408 & 0.5728 & \textbf{0.5056} & 0.7123 & 0.1190 & 29.3 & 242 &37\\

& {\color[HTML]{FF0000} X} & {\color[HTML]{FF0000} X}                        & {\color[HTML]{00CD66} Y}                       & \revision{$C_1$}    & \revision{\textbf{0.2943}} & \revision{\underline{0.7431}} & \revision{0.5685} & \revision{\underline{0.5018}} & \revision{0.7191} & \revision{0.1256} & \revision{28.1} & \revision{259} & \revision{34}\\

& {\color[HTML]{FF0000} X} & {\color[HTML]{00CD66} Y}                        & {\color[HTML]{00CD66} Y}                       & $C_2$    & 0.2247 & 0.7114 & \underline{0.5924} & 0.4716 & \underline{0.7208} & 0.1187 & \textbf{20.1} & 142 & \underline{33}\\
& {\color[HTML]{00CD66} Y}                        & {\color[HTML]{FF0000} X} & {\color[HTML]{00CD66} Y}                        & $C_3$    & 0.2277 & 0.6839 & 0.5685 & 0.4813 & 0.6690 & \underline{0.1262} & 20.6 & \underline{138} & 34 \\
&{\color[HTML]{00CD66} Y}                        & {\color[HTML]{00CD66} Y}                        & {\color[HTML]{FF0000} X} & $C_4$    & 0.2454 & 0.7002 & 0.5766 & 0.4772 & 0.6862 & 0.0890 & 23.6 & \textbf{136} & 45 \\
& {\color[HTML]{00CD66} Y}                        & {\color[HTML]{00CD66} Y}                        & {\color[HTML]{00CD66} Y}                        & $C_5$    & 0.2718 & \textbf{0.7445} & \textbf{0.6172} & 0.4909 & \textbf{0.7393} & \textbf{0.1338} & \underline{20.4} & 141 & \textbf{32} \\ \bottomrule
\end{tabular}}
\end{table}


\begin{tcolorbox}[colback=gray!20, colframe=gray!20, width=\columnwidth, left=0.05in, right=0.05in, top=0.05in, bottom=0.05in]
\revision{\textbf{Answer to RQ3:} 
Each component of \scheme\ contributes positively to performance.Input compression streamlines visual inputs, while ADTS suppresses repetition; removing either component degrades both efficiency and accuracy, indicating their complementarity. Consequently, the full model achieves the best results, outperforming any subset of
components.}
\end{tcolorbox}

\begin{table*}[ht]
\centering
\setlength{\tabcolsep}{0.1em}
\caption{\revision{Performance Comparison on WebCode2M. The compression ratio is} \textcolor{green!50!black}{\revision{63.42\%}}\revision{. Bold values indicate the optimal performance, and underlined values indicate the second best performance. RS is the number of samples with redundant codes.}}
\label{tab:RQ4_qwen}
\resizebox{0.8\textwidth}{!}{
\begin{tabular}{l|cccccccccc}
\hline
\multirow{2}{*}{\revision{Method}} & \multicolumn{10}{c}{\textbf{\revision{Webcode2M}}} \\
\cline{2-11}
& \revision{Block} & \revision{Text} & \revision{Position} & \revision{Color} & \revision{CLIP} & \revision{BLEU} & \revision{RS} & \revision{FLOPs(T)} & \revision{PT(ms)} & \revision{IT(s)} \\
\hline
\multicolumn{11}{c}{\it \revision{Qwen2.5-VL-32B}} \\
\hline
\revision{Vanilla}          & \textbf{\revision{0.8683}} & \underline{\revision{0.9621}} & \revision{0.7996} & \revision{0.7544} & \revision{0.8624} & \underline{\revision{0.6950}} & \revision{8} & \revision{84.5} & \revision{976} & \revision{92} \\
\revision{Random}           & \revision{0.6653} & \revision{0.9001} & \underline{\revision{0.8158}} & \textbf{\revision{0.7987}} & \revision{0.7967} & \revision{0.4086} & \underline{\revision{3}} & \revision{35.4} & \underline{\revision{510}} & \revision{80} \\
\revision{Visionzip}        & \revision{0.6283} & \revision{0.9054} & \revision{0.8155} & \revision{0.7718} & \underline{\revision{0.8715}} & \revision{0.4849} & \revision{5} & \underline{\revision{34.6}} & \revision{585} & \underline{\revision{77}} \\
\revision{EfficientUICoder} & \underline{\revision{0.8661}} & \textbf{\revision{0.9692}} & \textbf{\revision{0.8410}} & \underline{\revision{0.7932}} & \textbf{\revision{0.8802}} & \textbf{\revision{0.6999}} & \textbf{\revision{1}} & \textbf{\revision{34.5}} & \textbf{\revision{492}} & \textbf{\revision{74}} \\
\hline
\end{tabular}}
\end{table*}

\subsection{Parameter Study (RQ4)}



We sample 100 webpages in Design2Code and 50 webpages in WebCode2M for parameter study.

\subsubsection{\revision{Other Backbones}}

\revision{We additionally evaluate EfficientUICoder using Qwen2.5-VL Series~\cite{Qwen2.5-VL}, another state-of-the-art open-source MLLM. Since some baselines (fasv and pdrop) lack Qwen-based implementations, we compare our model against vanilla, random and visionzip. We report Qwen2.5-VL-32B's performance on WebCode2M in Table~\ref{tab:RQ4_qwen}, indicating that \textbf{EfficientUICoder achieves the best overall webpage quality and efficiency on Qwen2.5-VL-32B, further demonstrating EfficientUICoder's generalizability.}}


\subsubsection{The Suppression Step $s$ and Decay Factor $\lambda$} 
\begin{figure}[t]
    \subfigure[Block Match.]{
    \centering
    \includegraphics[width = .23\textwidth]{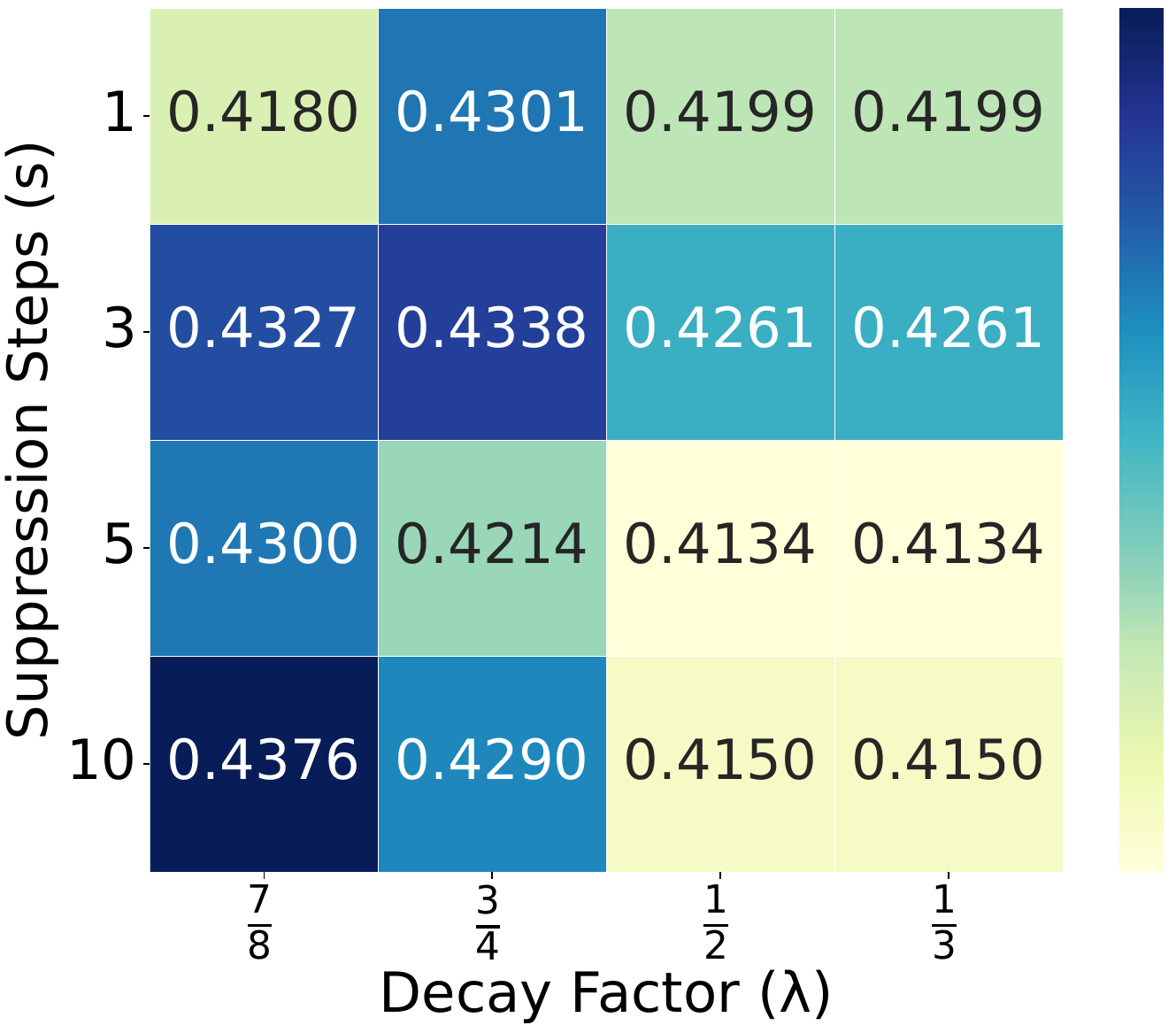}
    }    
    \subfigure[Text Match.]{
    \centering
    \includegraphics[width = .23\textwidth]{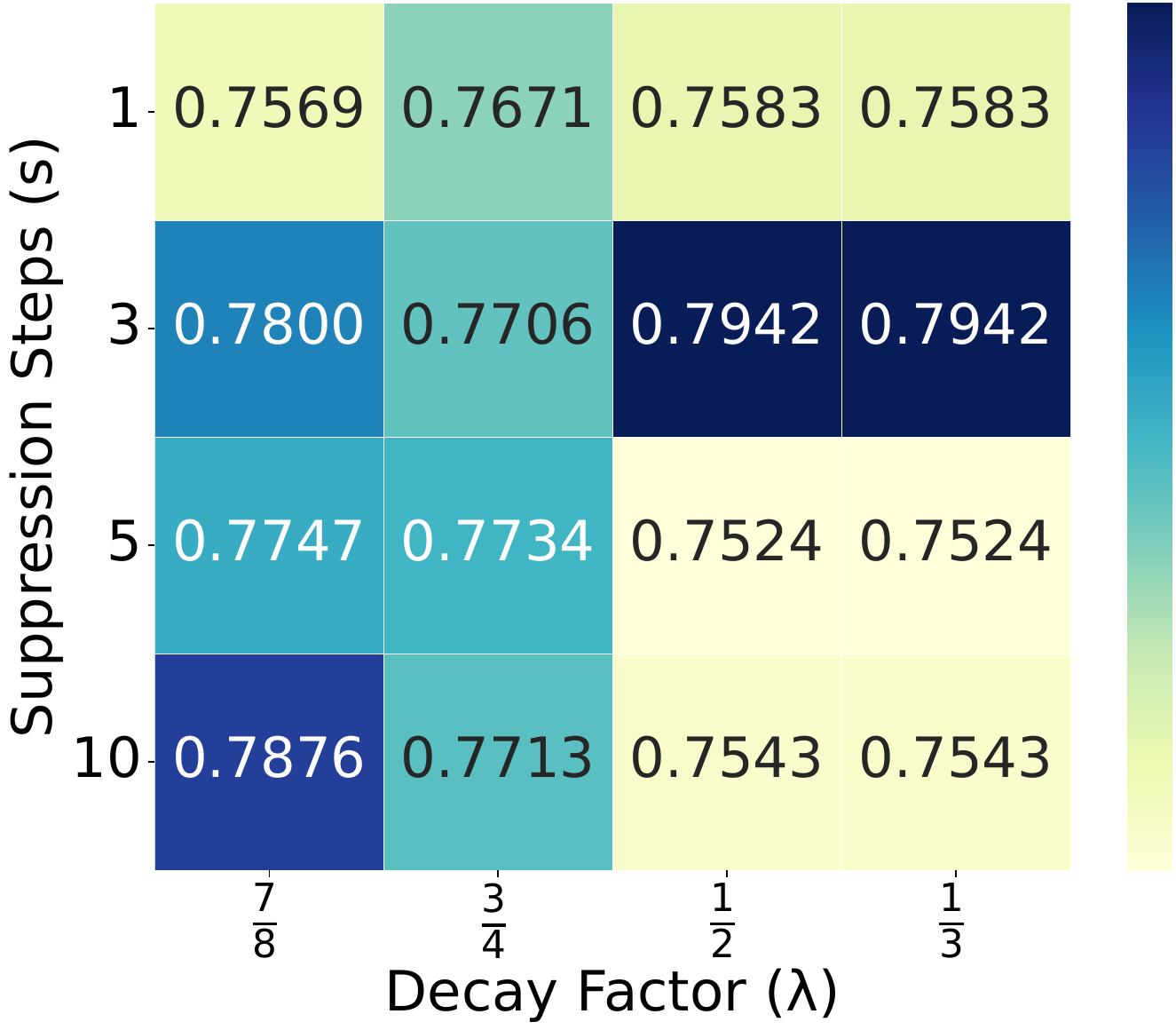}
    }
    \subfigure[CLIP Score.]{
    \centering
    \includegraphics[width = .23\textwidth]{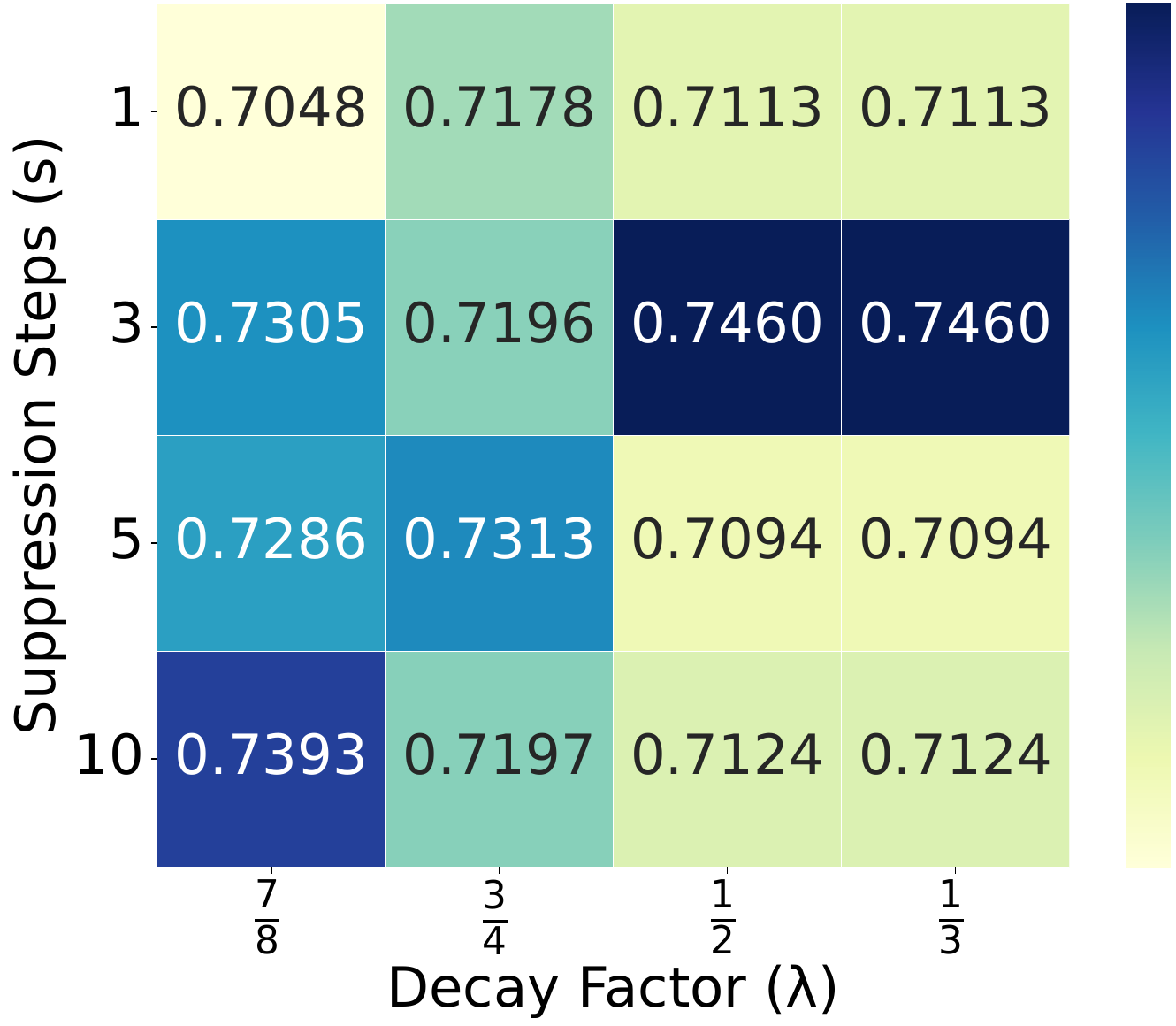}
    }
    \subfigure[Bleu.]{
    \centering
    \includegraphics[width = .23\textwidth]{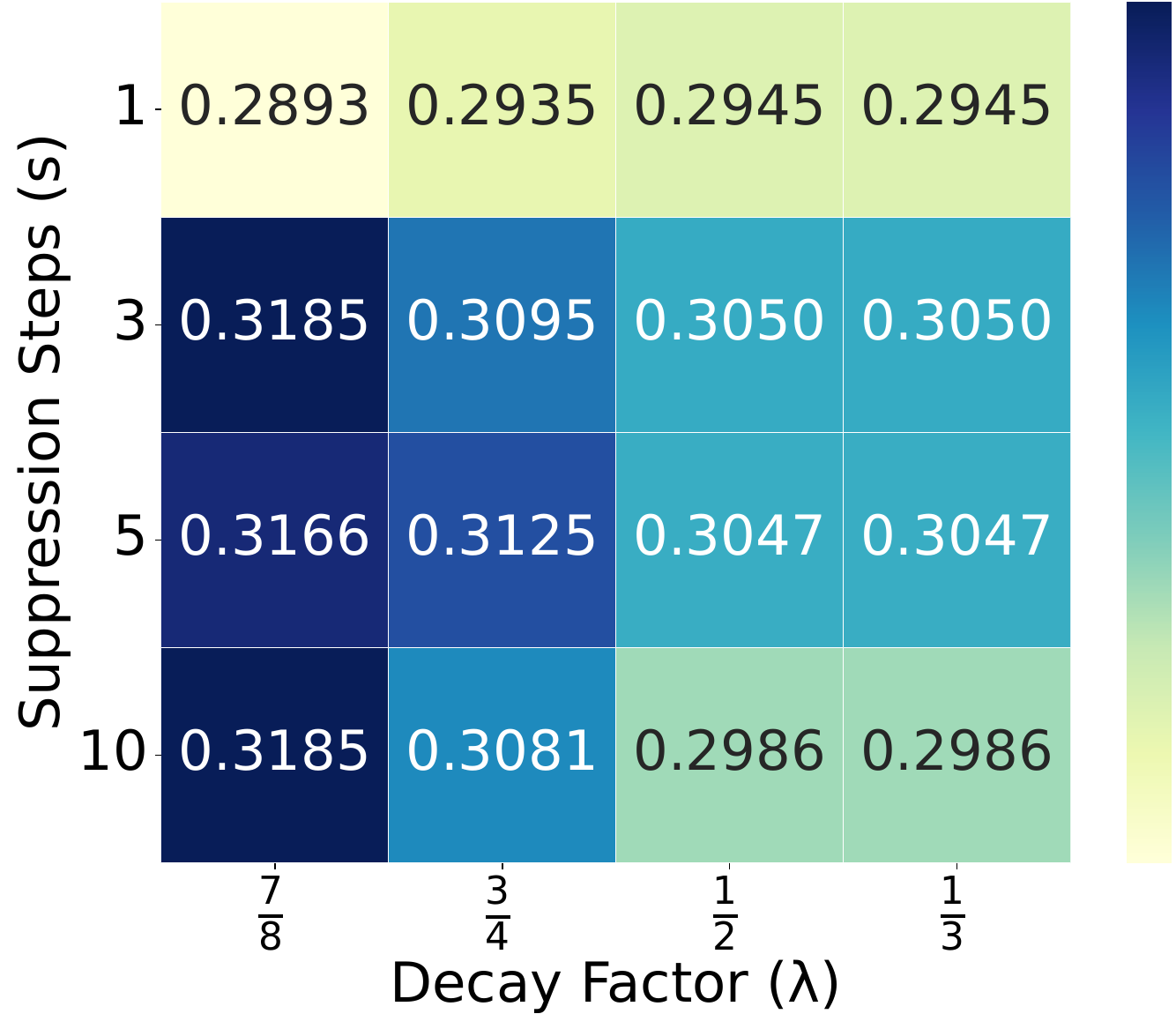}
    }
    \caption{Suppression step $s$ and decay factor $\lambda$ analysis on Design2Code \revision{dataset}.}
    \label{fig:d2c_para_performance}
\end{figure}

\begin{figure}[t]
    \subfigure[Block Match.]{
    \centering
    \includegraphics[width = .23\textwidth]{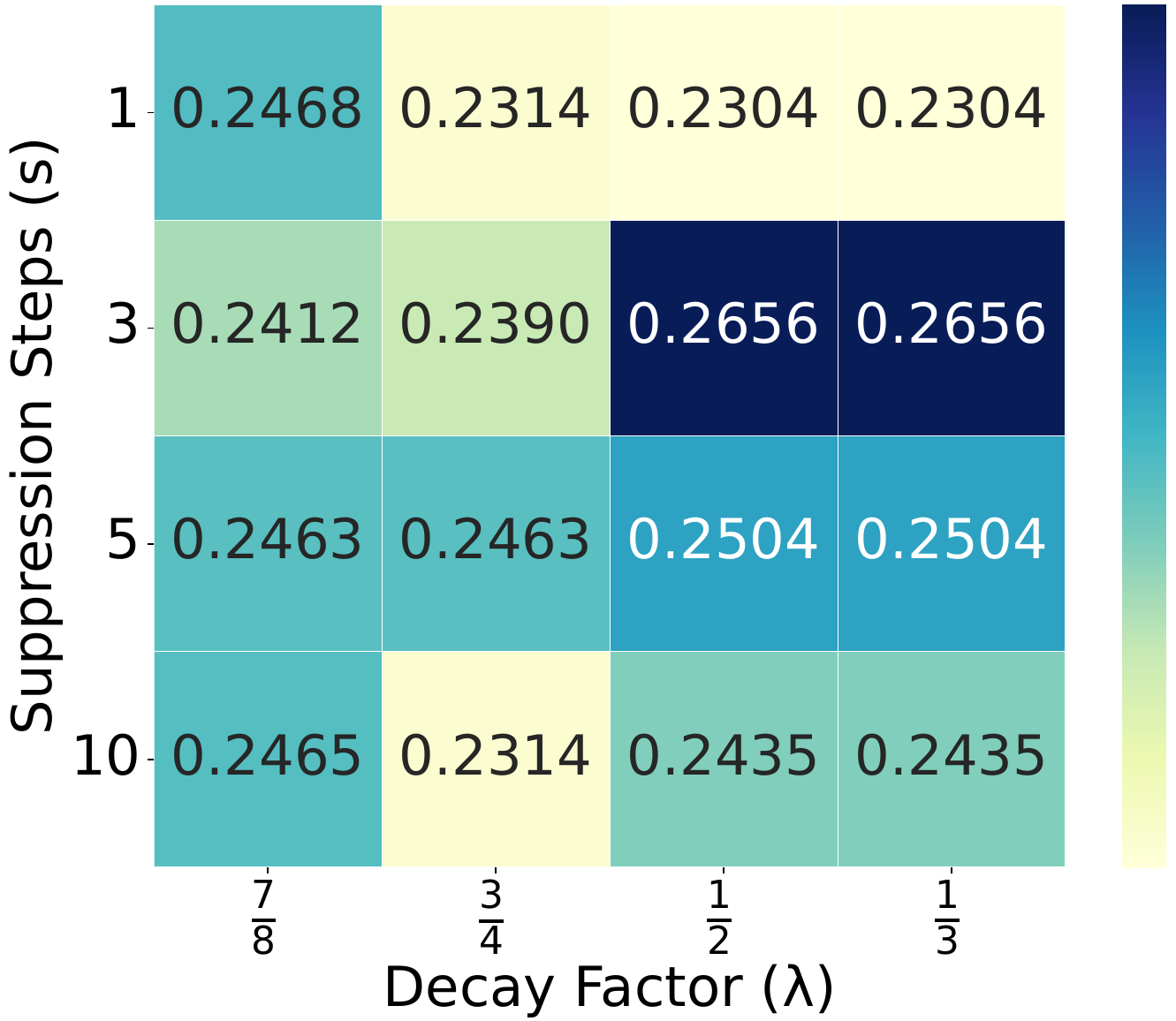}
    }    
    \subfigure[Text Match.]{
    \centering
    \includegraphics[width = .23\textwidth]{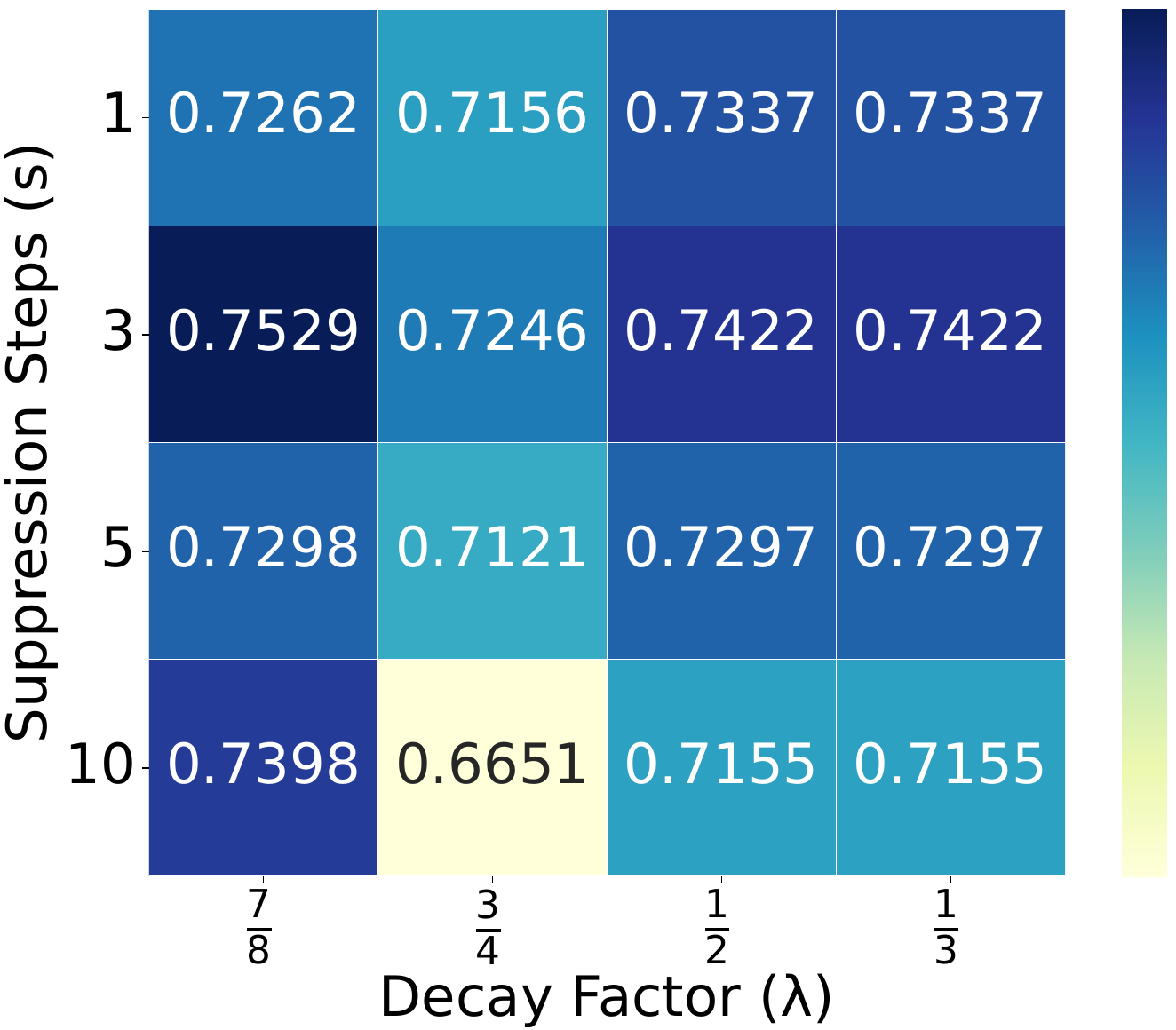}
    }
    \subfigure[CLIP Score.]{
    \centering
    \includegraphics[width = .23\textwidth]{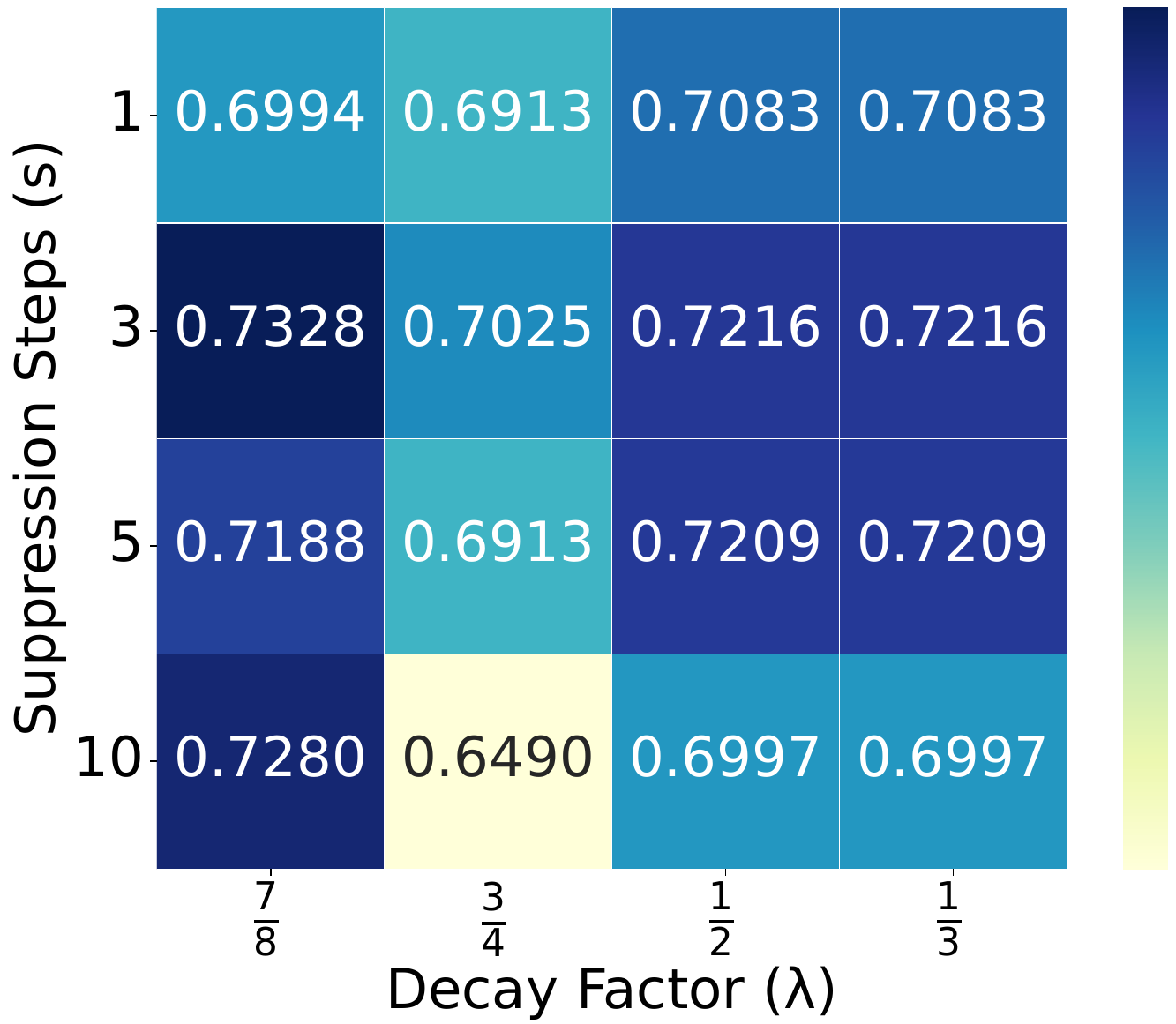}
    }
    \subfigure[Bleu.]{
    \label{fig:dis2}
    \centering
    \includegraphics[width = .23\textwidth]{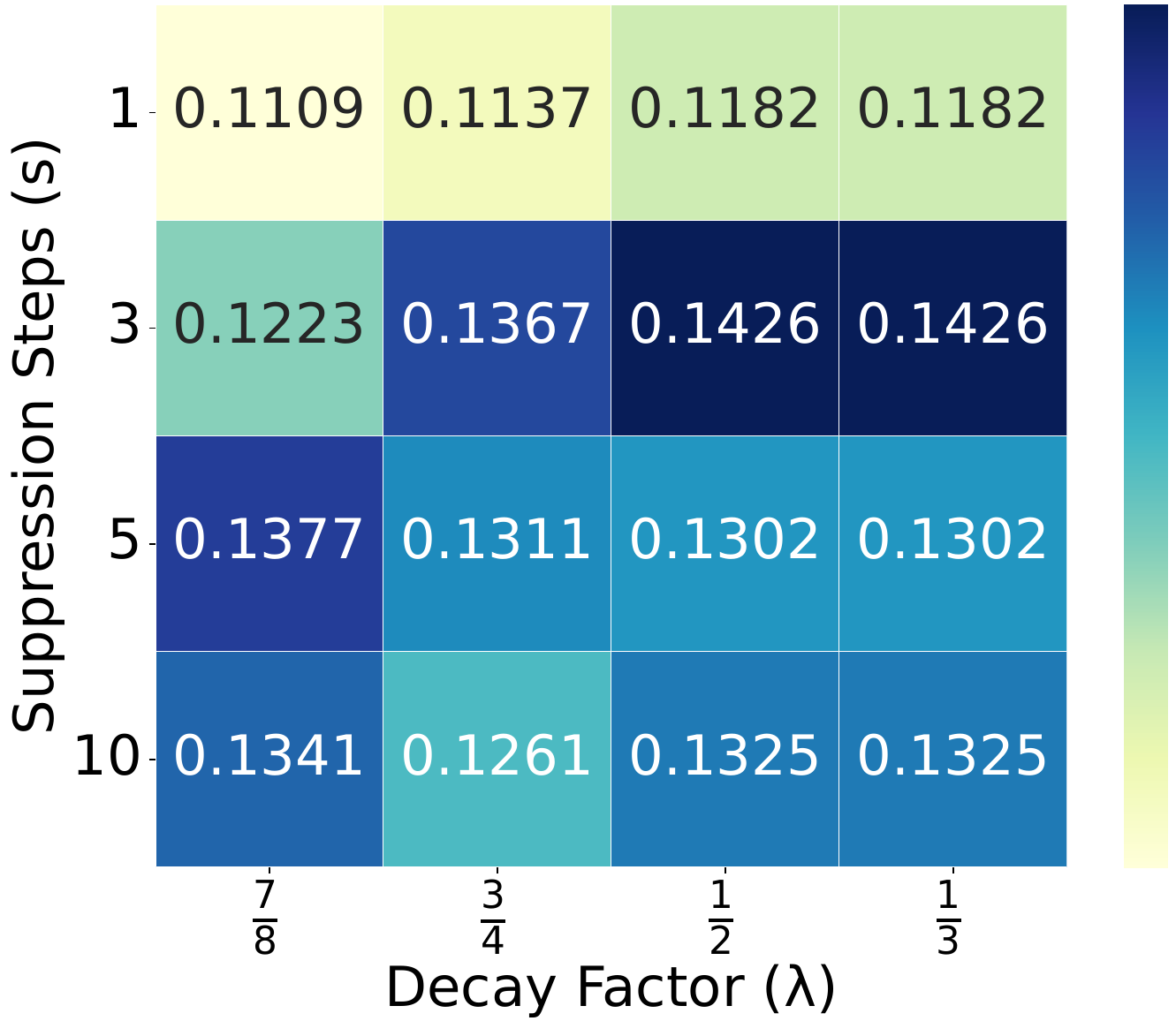}
    }
    \caption{Suppression step $s$ and decay factor $\lambda$ analysis on WebCode2M \revision{dataset}.}
    \label{fig:webcode2m_para_performance}
\end{figure}

We conduct a comprehensive parameter study by setting $s \in [1, 3, 5, 10]$ and $\lambda \in [\frac{7}{8}, \frac{3}{4}, \frac{1}{2}, \frac{1}{3}]$. As shown in Fig.~\ref{fig:webcode2m_para_performance} and Fig.~\ref{fig:d2c_para_performance}, \scheme \ achieves optimal performance when $s=3$ and $\lambda=\frac{1}{2}$. A suppression step that is too small ($s<3$) fails to sufficiently penalize consecutive duplicate tokens, while an excessively large step ($s>3$) results in over-penalization. Similarly, when the decay factor $\lambda$ is too large (close to 1), the logits of redundant code are not effectively reduced, whereas smaller values ($\lambda \leq \frac{1}{2}$) achieve similar effects in encouraging non-redundant token selection. Fig.~\ref{fig:inference_time_para} demonstrates that when $s \geq 3$, inference time is significantly reduced to approximately 30 seconds. Our analysis reveals that larger decay factors should be paired with larger penalty steps, while smaller decay factors work better with smaller penalty steps, achieving an optimal balance between penalty intensity and frequency to avoid under- or over-penalization of redundant content.


\subsubsection{The Refinement Token Ratio $r$} 
\begin{figure}[t]
    \centering
    \begin{minipage}{0.48\linewidth}
        \centering
        \subfigure[Design2Code.]{
            \includegraphics[width=0.45\linewidth]{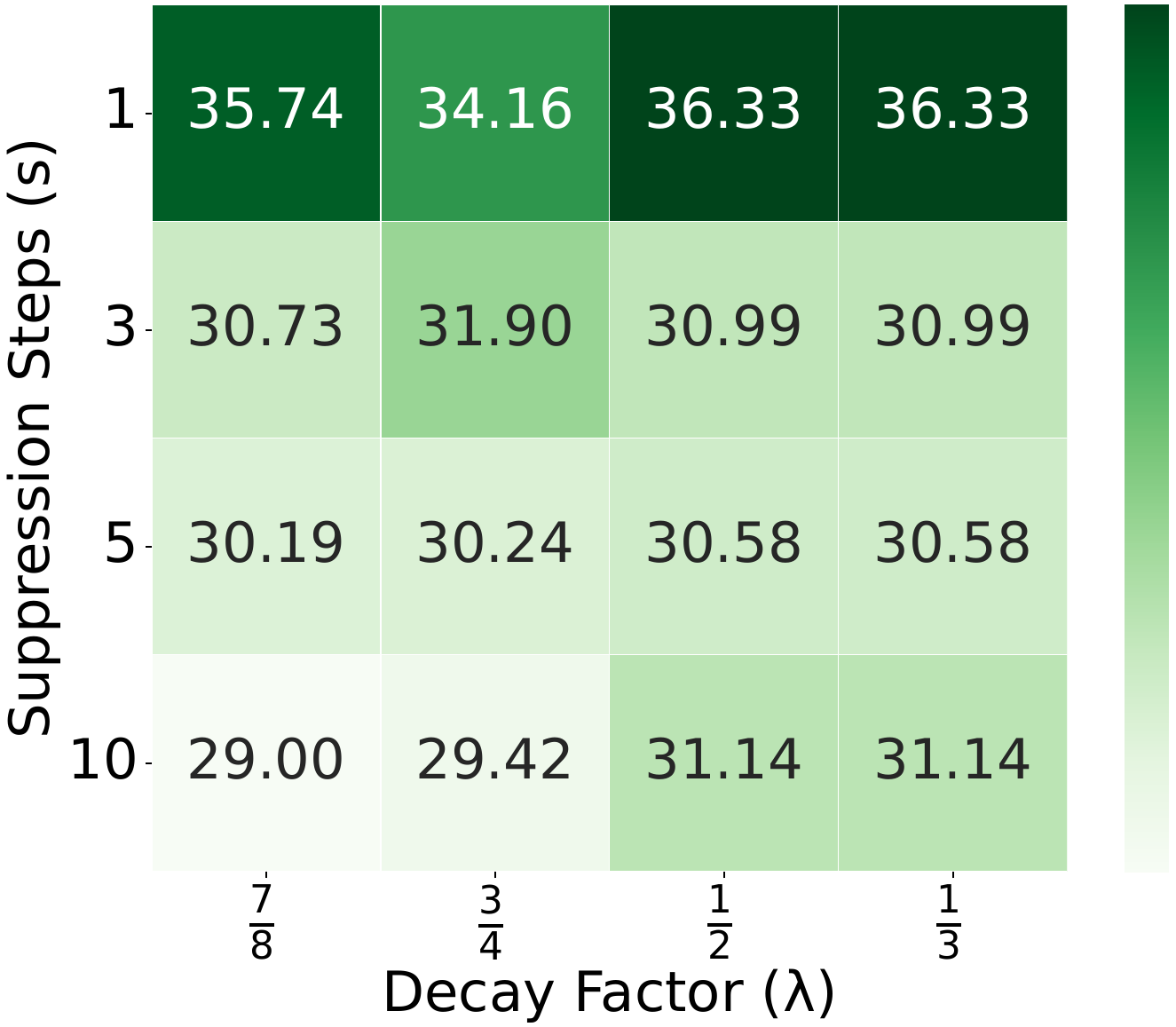}
        }
        \subfigure[WebCode2M.]{
            \includegraphics[width=0.45\linewidth]{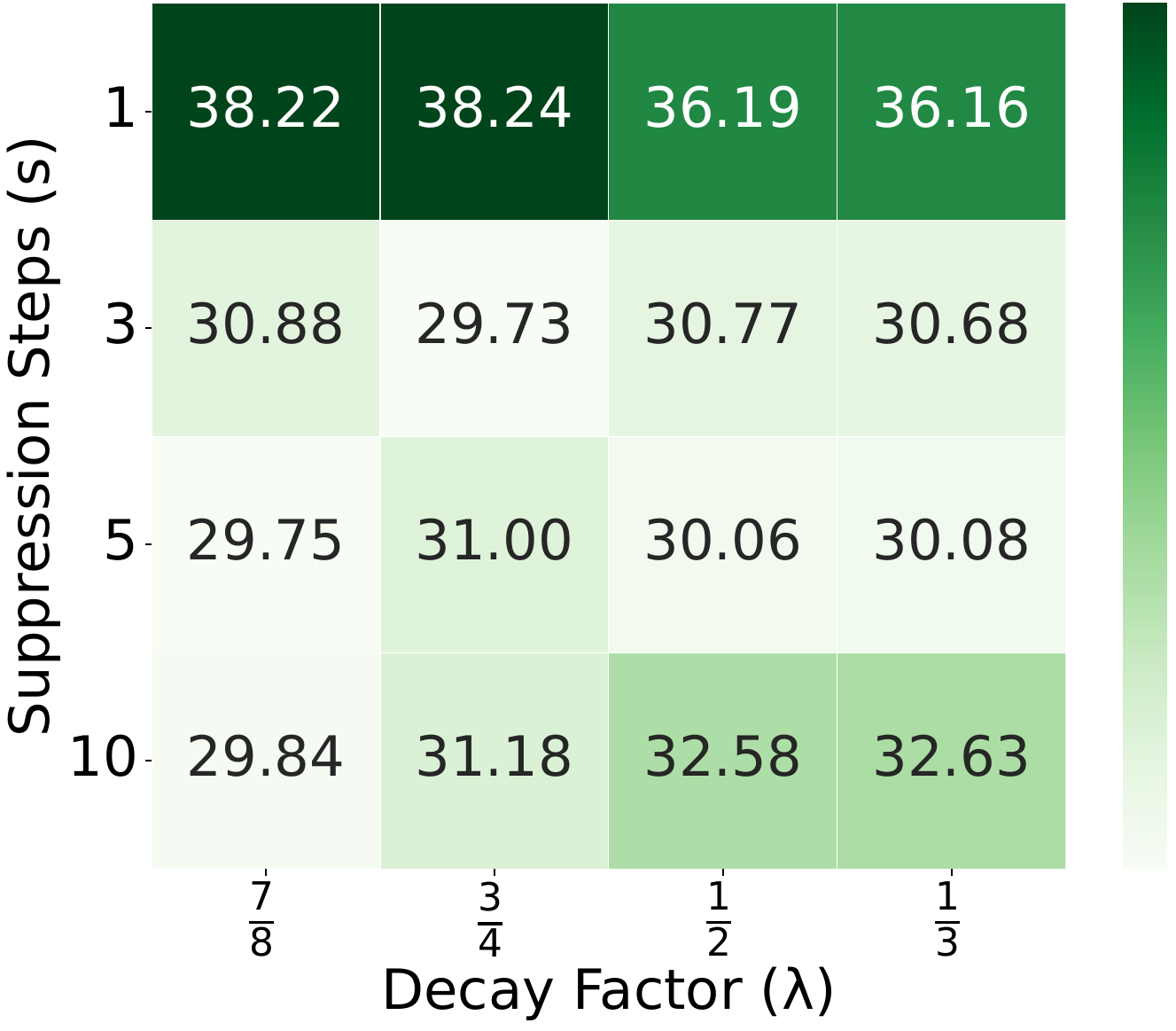}
        }
        \caption{Inference time under different $\lambda$ and $r$.}
        \label{fig:inference_time_para}
    \end{minipage}
    \hfill
    \begin{minipage}{0.48\linewidth}
        \centering
        \subfigure[Design2Code.]{
            \includegraphics[width=0.45\linewidth]{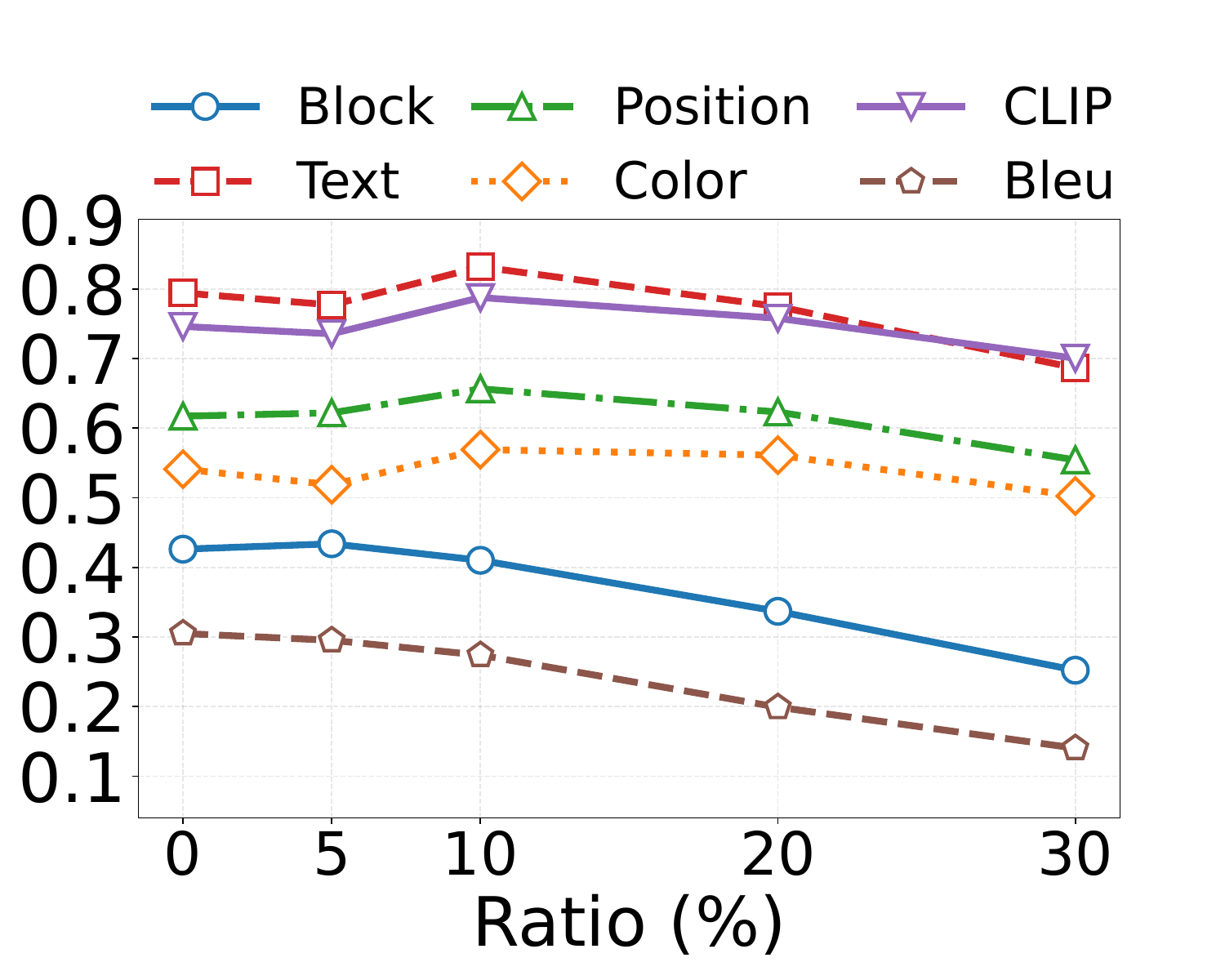}
        }
        \subfigure[WebCode2M.]{
            \includegraphics[width=0.45\linewidth]{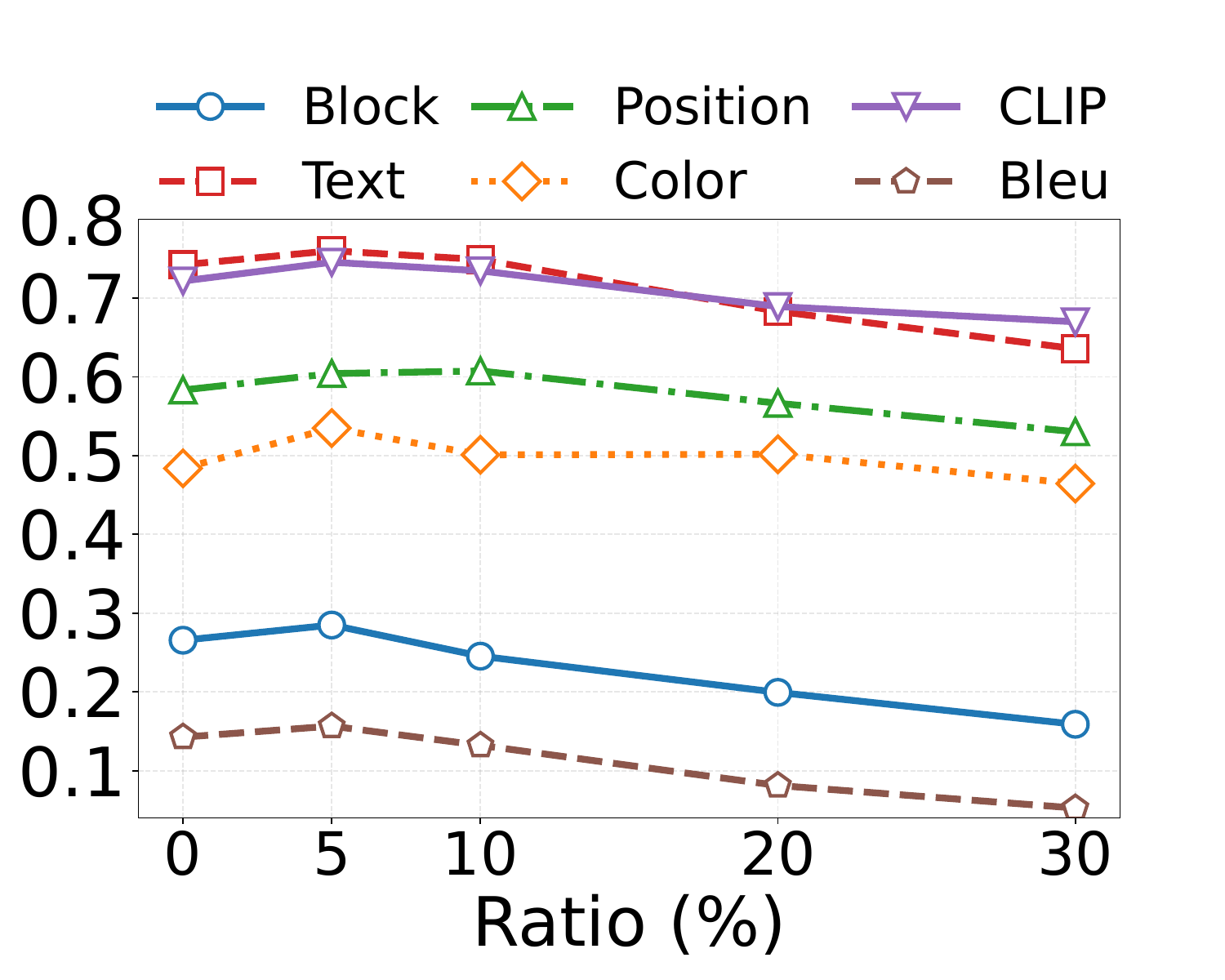}
        }
        \caption{Refinement ratio $r$ analysis.}
        \label{fig:refinement_ratio}
    \end{minipage}
\end{figure}

Maintaining the optimal parameter combination ($s=3$, $\lambda=\frac{1}{2}$), we investigate the influence of the refinement ratio by setting $r \in [5\%, 10\%, 20\%, 30\%]$. Fig.~\ref{fig:refinement_ratio} shows that \scheme \ achieves the best performance on Design2Code and WebCode2M datasets when $r=10\%$ and $r=5\%$, respectively. The performance exhibits an inverted-U shape as $r$ increases: initially improving as \scheme \ effectively discards redundant tokens in the ELTC region while incorporating important tokens from the non-ELTC region. However, excessively large values of $r$ lead to performance degradation due to the inadvertent removal of key tokens within the ELTC region, which represents crucial UI elements and layout information.


\begin{figure}[t]
    \subfigure[Performance on D2C.]{
    \centering
    \includegraphics[width = .21\textwidth]{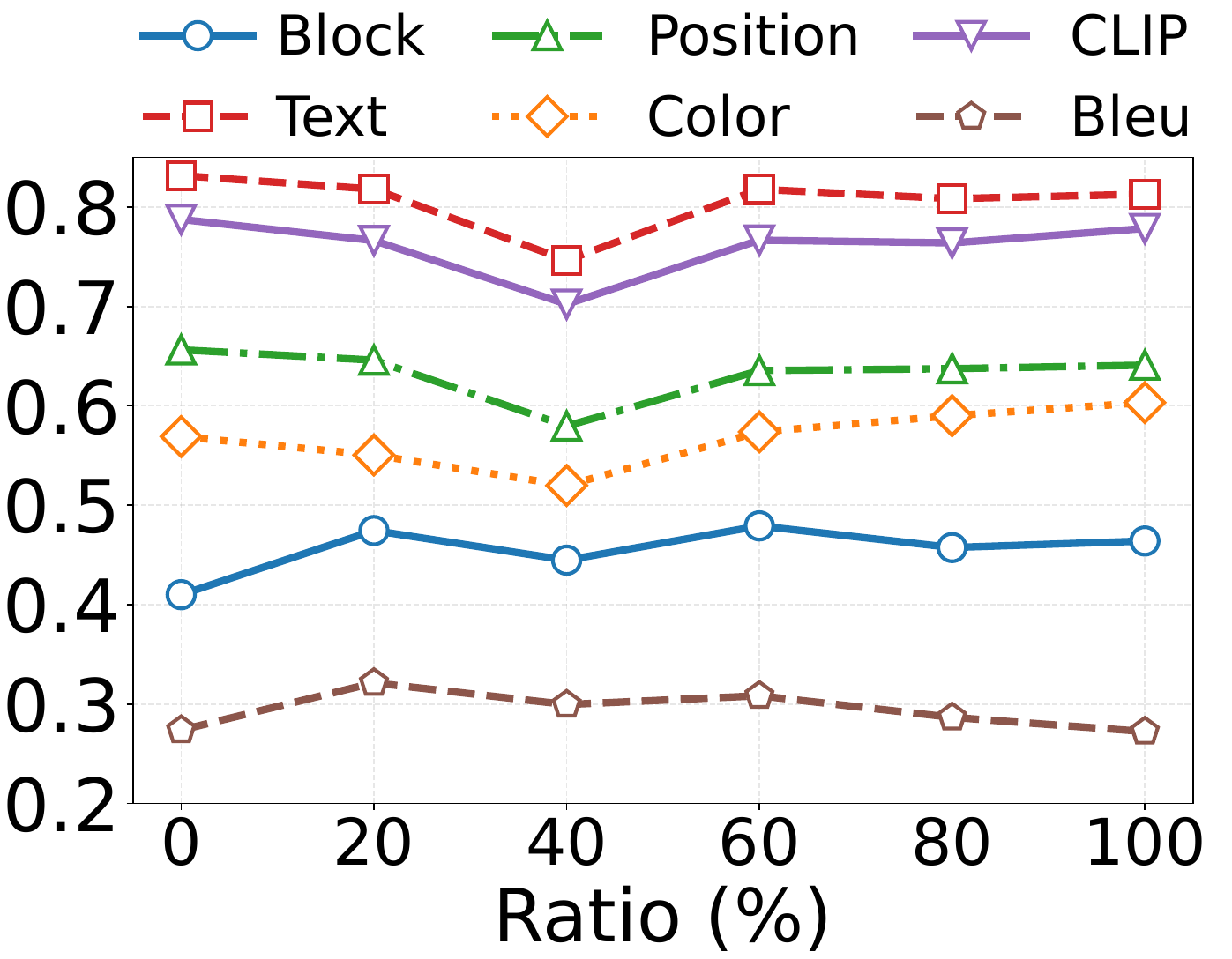}
    }    
    \subfigure[Performance on WC2M.]{
    \centering
    \includegraphics[width = .21\textwidth]{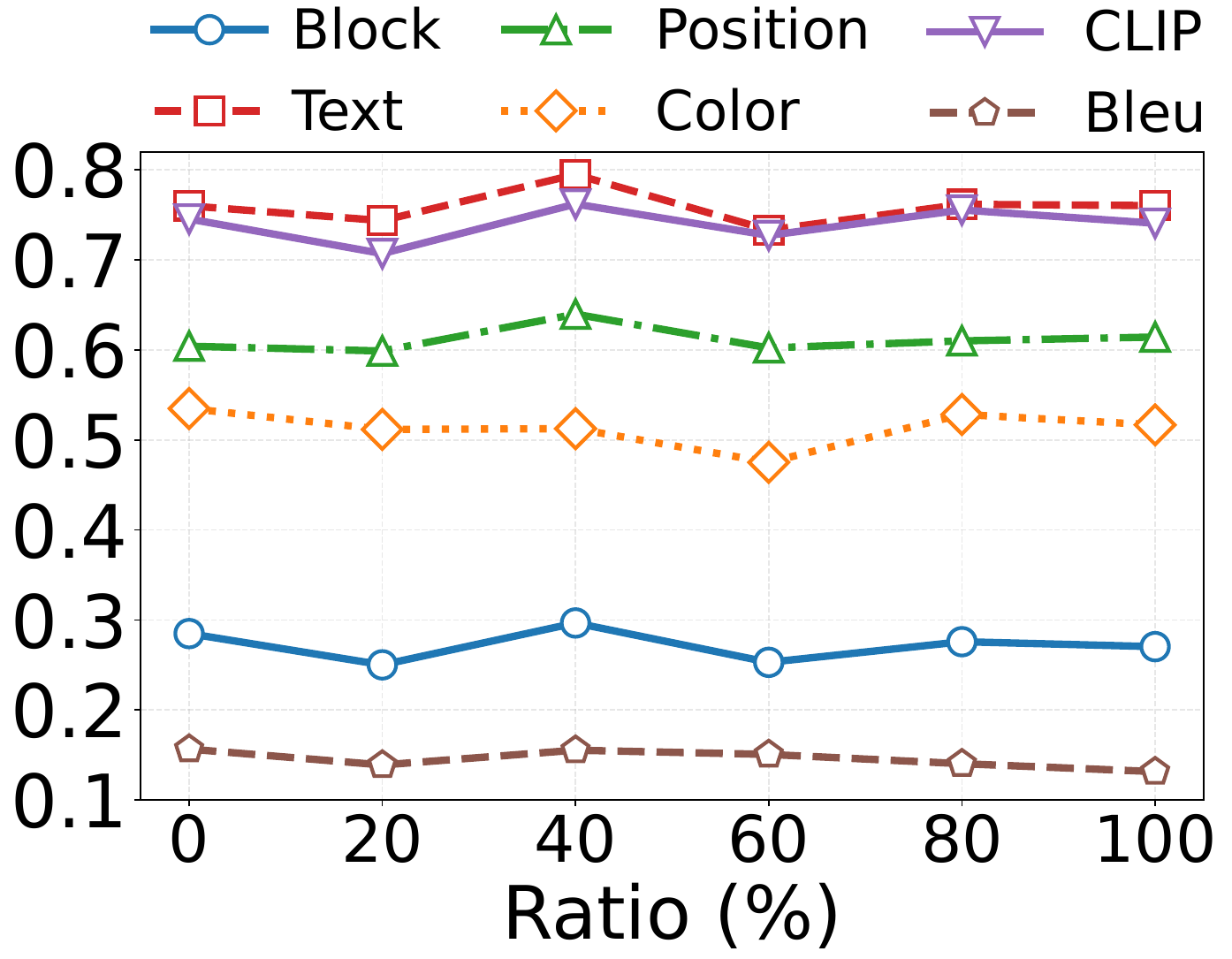}
    }
    \subfigure[Efficiency on D2C.]{
    \centering
    \includegraphics[width = .25\textwidth]{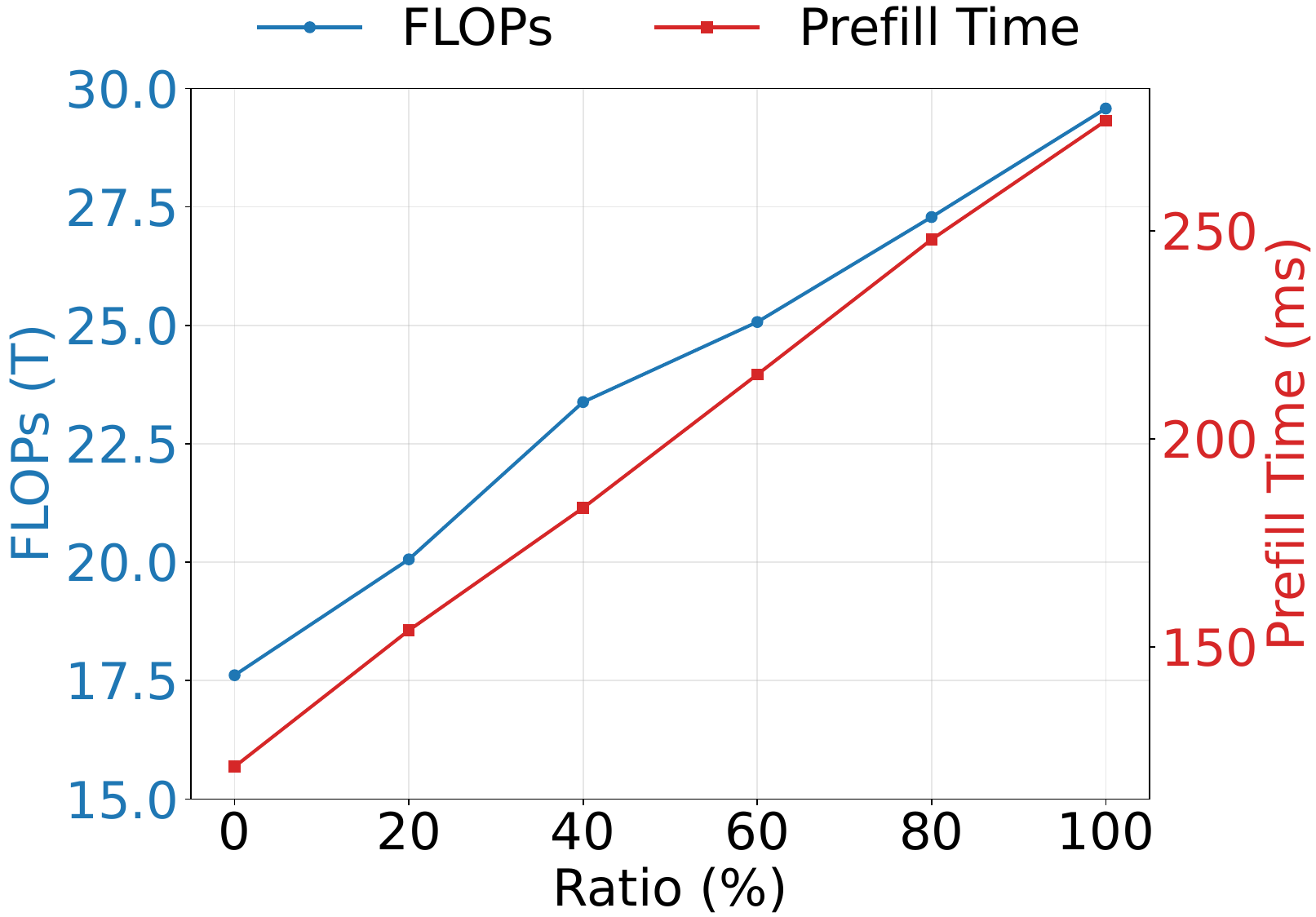}
    }
    \subfigure[Efficiency on WC2M.]{
    \centering
    \includegraphics[width = .25\textwidth]{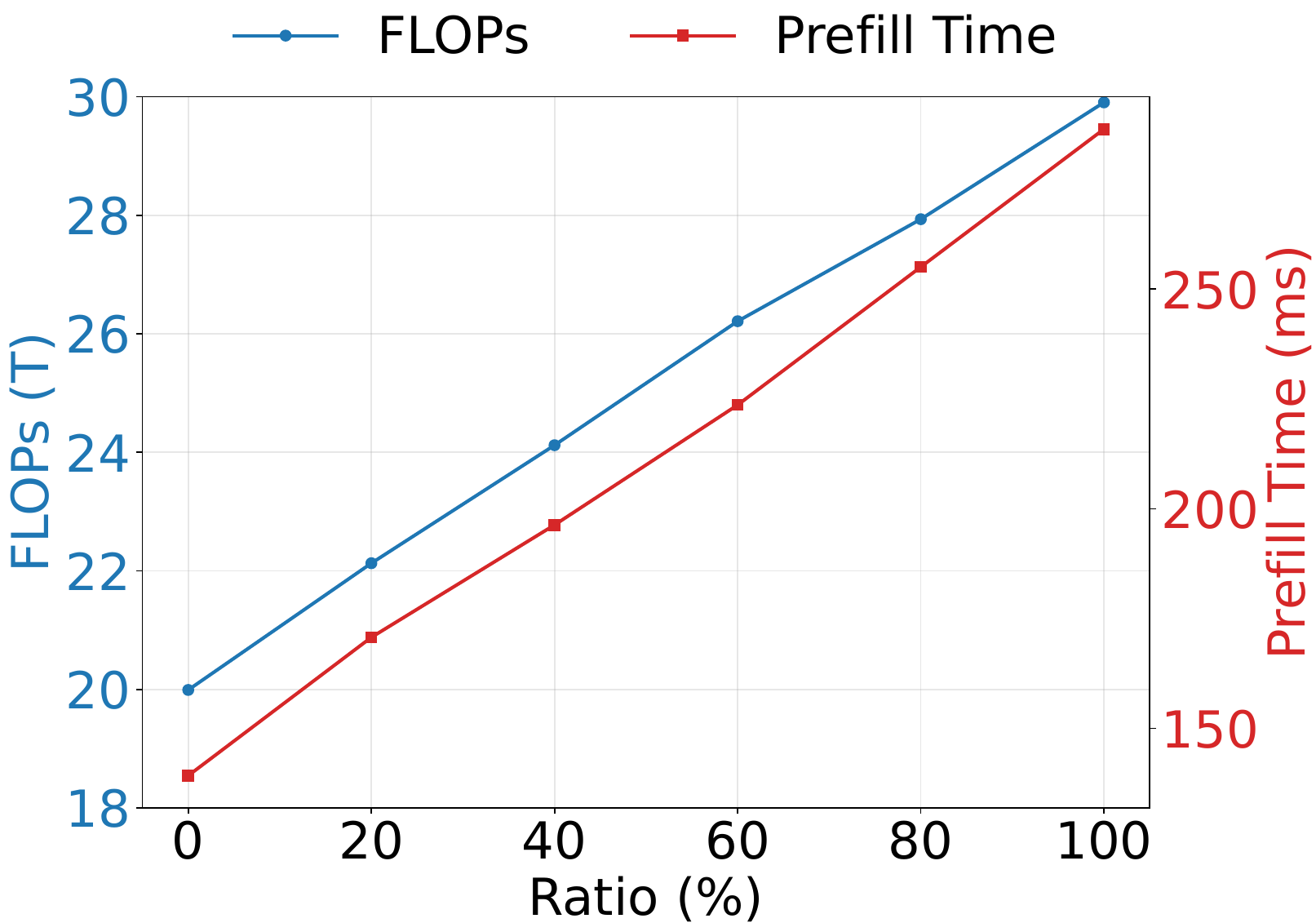}
    }
    \caption{The performance and efficiency on two datasets when adding unselected token. D2C denotes Design2Code and WC2M denotes WebCode2M.}
    \label{fig:token_add}
\end{figure}


\subsubsection{The Proportion of Increased Tokens $a$} To further validate the effectiveness of our token selection strategy, we explore the impact of adding a certain proportion of tokens from the initially unselected set. We divide the unselected set into five parts, and add $\frac{1}{5}$ of the tokens each time ($a\in[20\%, 40\%, 60\%, 80\%, 100\%]$). Fig.~\ref{fig:token_add} reveals that increasing $a$ does not yield significant performance improvements but instead causes fluctuations. This observation confirms the effectiveness of \scheme's token selection strategy, demonstrating that adding more tokens provides only marginal performance gains while substantially increasing computational costs.



\begin{tcolorbox}[colback=gray!20, colframe=gray!20, width=\columnwidth, left=0.05in, right=0.05in, top=0.05in, bottom=0.05in]
\textbf{Answer to RQ4:} 
A moderate suppression step $s=3$ and decay factor $\lambda=\frac{1}{2}$ achieve a good balance between under- and over-penalization. The refinement ratio $r$ works best at small values (5-10\%), effectively removing redundancy without discarding key UI elements. \revision{\scheme \ shows  generalizability across different MLLM backbones.}
\end{tcolorbox}

\subsection{Case Study (RQ5)}

\begin{figure}[t]
\includegraphics[width = .98\textwidth]{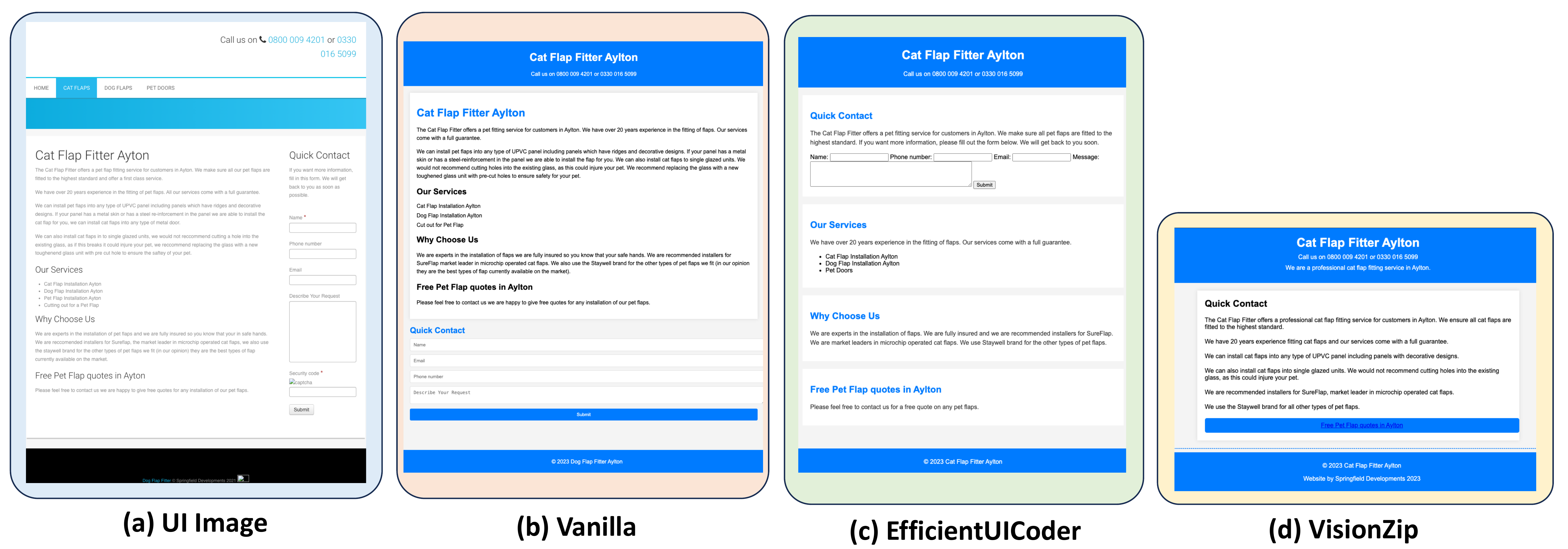}
\caption{Case study of webpages generated by Vanilla, \scheme \  and VisionZip.}
\label{fig:case1}
\end{figure}

\begin{figure}[t]
    \centering
    \begin{minipage}{0.44\linewidth}
        \centering
            \includegraphics[width=0.95\linewidth]{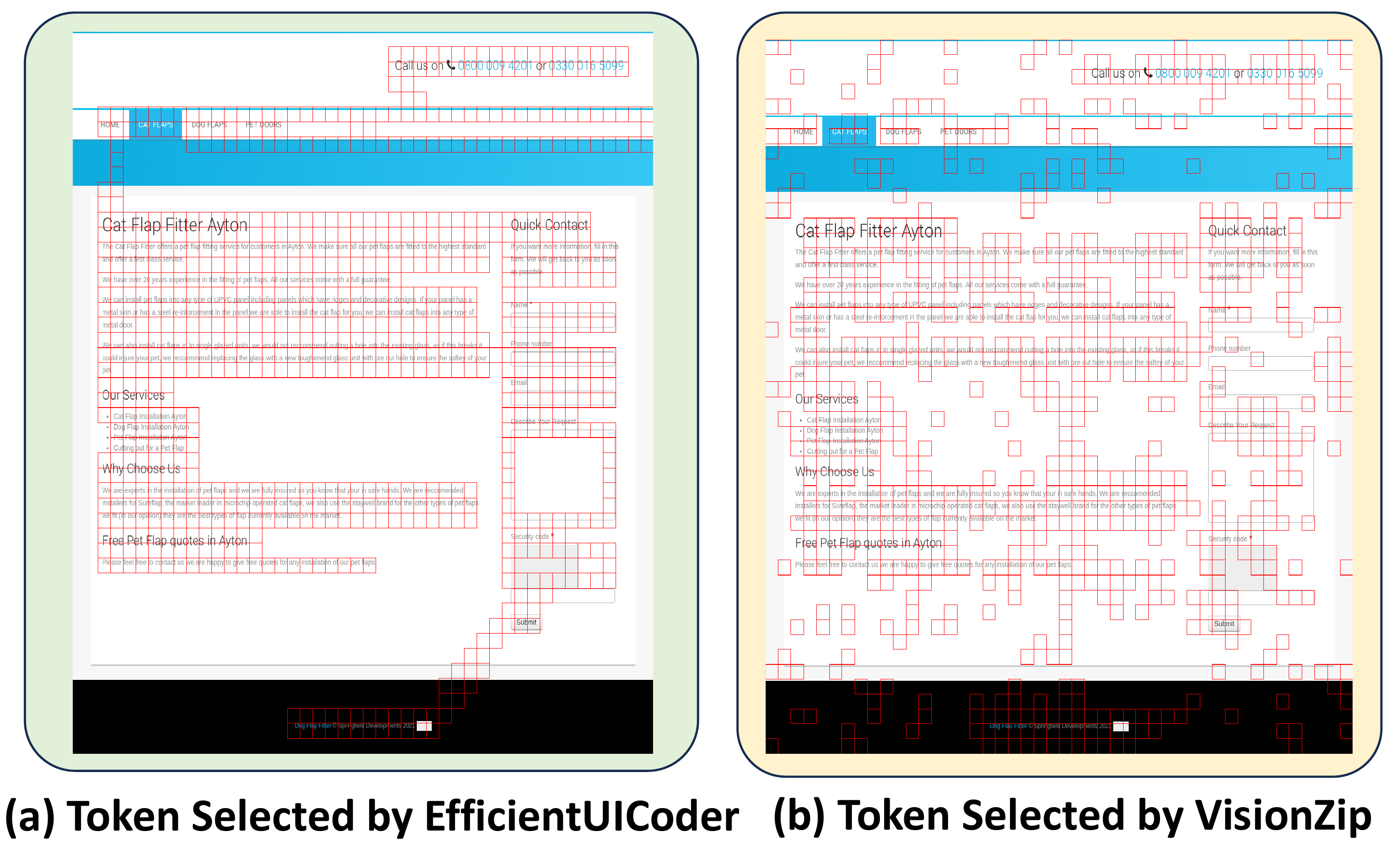}
        \caption{Token selection results.}
        \label{fig:case2}
    \end{minipage}
    \begin{minipage}{0.54\linewidth}
        \centering
            \includegraphics[width=0.99\linewidth]{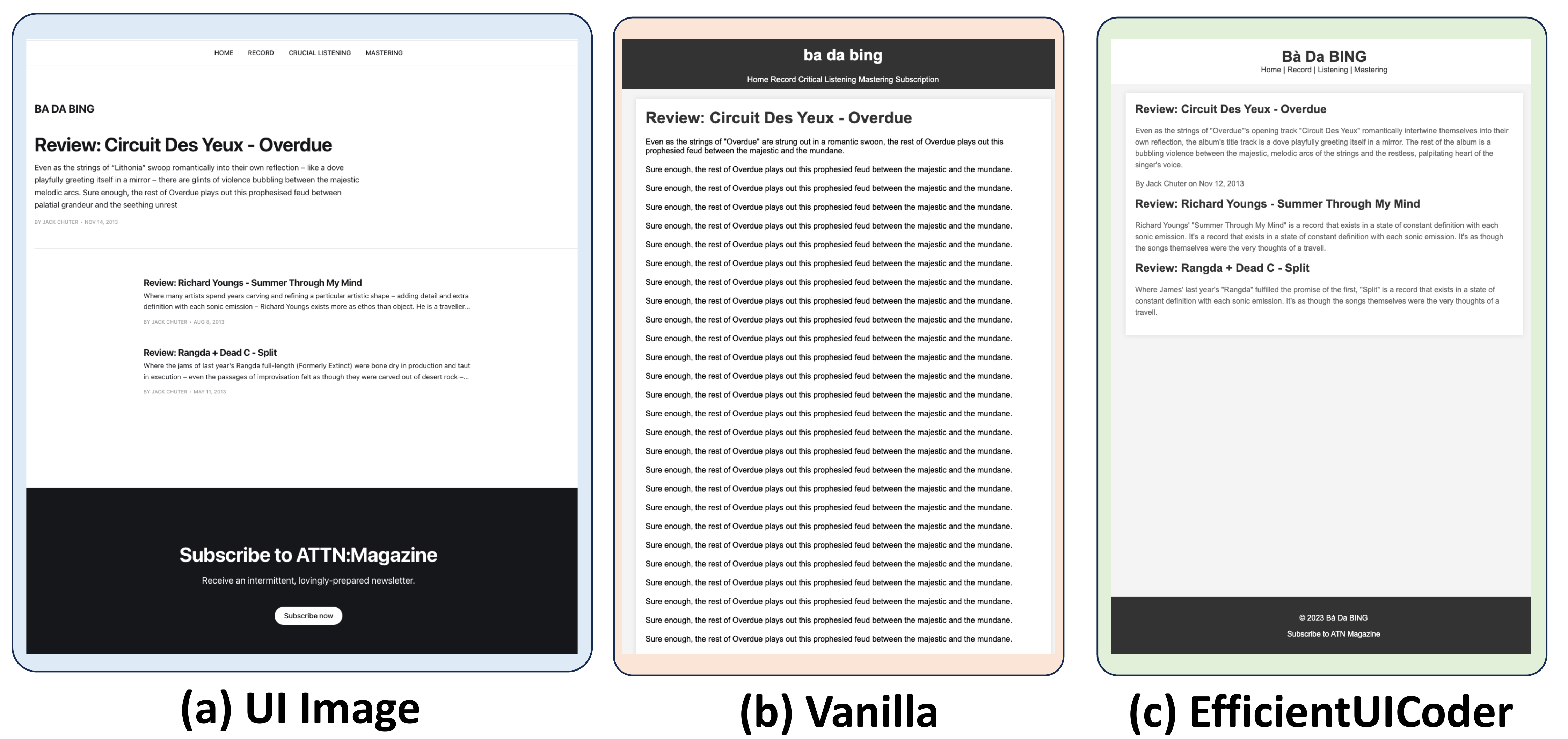}
        \caption{Output redundancy case study.}
        \label{fig:case3}
    \end{minipage}
\end{figure}

We present two cases generated by Llava-v1.6-34b from the Design2Code dataset with visualized token selection results to demonstrate the effectiveness of \scheme. (1) \scheme \ effectively selects tokens that encompass critical UI elements and structural components. As shown in Fig.\ref{fig:case1}, the webpage generated by \scheme achieves quality comparable to the Vanilla approach while significantly outperforming VisionZip. Detailed analysis reveals that VisionZip fails to generate essential elements including contact information (name, phone number, email), interactive components (message input boxes), and key section headers such as "Our Services" and "Why Choose Us". Fig.\ref{fig:case2} illustrates the token selection comparison: \scheme's selections cover nearly all UI elements, whereas VisionZip not only omits crucial text and component tokens but also includes numerous redundant background tokens. (2) \scheme \ effectively prevents the generation of duplicate code structures. Fig.~\ref{fig:case3} demonstrates a scenario where the Vanilla method becomes trapped in continuously generating the paragraph element "Sure enough,...", resulting in excessive webpage length and preventing the generation of subsequent elements. In contrast, \scheme \ successfully avoids this repetitive generation through its ADTS module, ensuring balanced and complete code synthesis.




\begin{tcolorbox}[colback=gray!20, colframe=gray!20, width=\columnwidth, left=0.05in, right=0.05in, top=0.05in, bottom=0.05in]
\textbf{Answer to RQ5:} EfficientUICoder works through two key mechanisms: (1) Effective token selection that covers critical UI elements while filtering redundant tokens, and (2) ADTS's redundancy prevention that avoids repetitive code generation, ensuring valid webpage.
\end{tcolorbox}

\section{Threats to Validity}

(1) \textit{Selection of backbone models.} We adopt widely used MLLMs, LLaVA and Qwen, with 7B and 30B parameters as backbones. Commercial LLMs are excluded because their closed-source nature prevents modifications to the encoding and decoding processes. (2) \textit{Metric bias.} We use BLEU and CLIP scores to measure code similarity and visual consistency, along with fine-grained metrics. Since no standardized UI2Code evaluation framework exists, these metrics may not fully capture task nuances; thus, we supplement them with human evaluation (Section~\ref{subsec:RQ1}).



\section{Related Work}

\textbf{UI Code Generation Method}. Early UI code generation methods relied on deep learning and computer vision techniques, such as CNN-based encoder-decoder frameworks~\cite{acsirouglu2019automatic,beltramelli2018pix2code,chen2022code} and OCR/object-detection-based systems~\cite{jain2019sketch2code,nguyen2015reverse}. Recent studies have shifted toward MLLM-based approaches to improve layout fidelity and element completeness. DCGen~\cite{wan2024automatically} adopts a divide-and-conquer strategy, DeclarUI~\cite{zhou2024bridging} integrates segmentation with transition graphs, and UICopilot~\cite{gui2025UICopilot}, LayoutCoder~\cite{wu2024mllm} and LatCoder~\cite{gui2025latcoder} further enhance generation through hierarchical and layout-aware modeling. TDDev~\cite{wan2025automatically} introduces a test-driven development to enable the generation of full-stack web applications. ComUICoder~\citep{xiao2026comuicoder} applies semantic-aware segmentation and merging techniques for component-based UI code generation, improving code reusability. However, these methods mainly focus on generation quality, while overlooking the computational overhead challenge.

\textbf{UI Code Generation Benchmark}. Early works like WebSight and Web2Code~\cite{yun2024web2code} pioneered HTML code synthesis and systematic assessment through the Webpage Code Generation Benchmark (WCGB), though both relied on synthetic data. Design2Code~\cite{si2025design2code} introduced the first real-world benchmark with 484 manually curated web pages from Common Crawl, while WebCode2M~\cite{gui2025webcode2m} scaled this to 20,000 samples for comprehensive training and evaluation. Specialized benchmarks have emerged targeting specific aspects: Interaction2Code~\cite{xiao2024interaction2code} for interactive generation, MRWeb~\cite{wan2024mrweb} for multi-page resource-aware websites, and DesignBench~\cite{xiao2025designbench} for multi-task framework-based UI generation, editing, and repair. These benchmarks collectively provide comprehensive evaluation frameworks for different dimensions of MLLM web development capabilities.

\textbf{Token Compression for MLLMs}. Existing token compression methods mainly reduce redundant visual tokens based on attention scores or token similarity. For example, FastV~\cite{chen2024fastv} drops half of visual tokens at the second layer during inference, Pdrop~\cite{xing2024pyramiddrop} performs stage-wise token pruning with predefined ratios, SparseVLM~\cite{zhang2024sparsevlm} leverages text-guided attention for adaptive pruning, and VisionZip~\cite{yang2025visionzip} filters tokens using visual encoder attention scores. However, these methods are not well suited for UI2Code tasks, as they ignore UI element semantics and do not address redundancy in generated outputs.

\revision{\textbf{Repetitive Content Compression.} Repetition penalty~\cite{holtzman2019curious, wolf2020huggingfacestransformers} is a widely adopted technique for mitigating content repetition during decoding by penalizing previously generated tokens. ARP~\cite{fu2021theoretical} proposes a rebalanced encoding strategy to alleviate repetitive content during the training phase. Repetition dropout~\cite{li2023repetition} reduces repetitive content by selectively dropping attention to repetitive tokens in the training data.} \revision{ADTS differs from above methods in four ways: (1) \textbf{Structure-aware counters}: instead of plain token strings tracking, we track <selector, property> tuples for CSS, 4-tuples <tag, attr, value, text> for HTML and substring for text content. (2) \textbf{Dynamic scope}: penalty is applied only inside the corresponding syntactic scope (style block or body), avoiding over-suppression of other contents. (3) \textbf{Exponential decay}: ADTS also incorporates an exponential penalty parameter, the more repetitions, the more pronounced the penalty, instead of completely suppressing it whenever repetition occurs. (4) \textbf{Lightweight}: ADTS does not require extensive training or the construction of large-scale datasets for fine-tuning.}








\section{Conclusion}
We present \scheme, a bidirectional compression framework for efficient UI code generation. We identify the substantial redundancies in both image and code tokens that not only increase computational overhead but also hinder model focus on key UI elements. To address these issues, \scheme \ integrates three complementary techniques: (i) Element and Layout-aware Token Compression, which condenses visual inputs while preserving essential UI structures; (ii) Region-aware Token Refinement, which selectively enhances semantically important tokens; and (iii) Adaptive Duplicate Token Suppression, which dynamically penalizes repetitive code generation. Experimental results demonstrate that \scheme \ achieves a 55\%--60\% compression ratio without sacrificing quality, while significantly reducing prefill time, FLOPS, and inference latency.

\section*{Data Availability}

All the code for \scheme \ are available at \url{https://github.com/WebPAI/EfficientUICoder}.

\section*{Acknowledgment}
The work was supported by three grants from the Research Grants Council of the Hong Kong Special Administrative Region, China: (1) No. SRFS2425-4S03 of the Senior Research Fellow Scheme, (2) No. CUHK 14209124 of the General Research Fund and (3) No. PF23-87650 of Jingyu Xiao’s Hong Kong PhD Fellowship Scheme. This research was also supported by the Singapore Ministry of Education (MOE) Academic Research Fund (AcRF) Tier 1 grant (Project ID: 23-SIS-SMU-088). Any opinions, findings and conclusions or recommendations expressed in this material are those of the author(s) and do not reflect the views of the Ministry of Education, Singapore.

\bibliographystyle{ACM-Reference-Format}
\bibliography{sample-base}


\end{document}